\documentclass[12pt]{article}
\usepackage{color}
\usepackage{amssymb,latexsym,bm,amsmath,amsthm,amsfonts,}
\usepackage{eufrak,booktabs}
\usepackage{multirow,graphicx,mathrsfs}
\usepackage{epsfig}
\usepackage{ulem}
\usepackage{enumitem}

\graphicspath{{figures/}}

\newcommand{\argmin}{\mathop{\rm argmin}}
\newcommand{\argmax}{\mathop{\rm argmax}}

\def\vec{{\rm vec}}

\def\1{{\bm 1}}
\def\0{{\bm 0}}
\def\th{{\rm th}}

\def\gic{{\rm GIC}}
\def\aic{{\rm AIC}}
\def\bic{{\rm BIC}}

\def\tr{{\rm tr}}

\def\IF{{\rm IF}}

\def\D{{\cal D}}

\def\I{\mathcal{I}}

\newcommand{\real}{\mathbb{R}}

\renewcommand{\hat}{\widehat}

\newtheorem{thm}{Theorem}
\newtheorem{defn}{Definition}

\newtheorem{rmk}{Remark}

\title{A generalized information criterion for high-dimensional PCA rank selection}

\begin{document}

\author{Hung Hung$^{1}$, Su-Yun Huang$^{2}$
and Ching-Kang Ing$^{3}$\\[2ex]
\small $^1$Institute of Epidemiology and Preventive Medicine,
National
Taiwan University, Taiwan\\
\small $^2$Institute of Statistical Science, Academia Sinica,
Taiwan\\
\small $^3$Institute of Statistics, National Tsing Hua University, Taiwan}
\date{}
\maketitle

\begin{abstract}
Principal component analysis (PCA) is a commonly used
statistical procedure for dimension reduction. An important issue
for PCA is to determine the rank, which is the number of
dominant eigenvalues of the covariance matrix. Among information-based criteria, Akaike information criterion (AIC) and
Bayesian information criterion (BIC) are two most common ones. Both use the
number of free parameters for assessing model complexity, which may suffer the problem of model misspecification. To alleviate this difficulty, we propose using the generalized information criterion (GIC) for PCA rank selection. The resulting GIC model complexity takes into
account the sizes of eigenvalues and, hence, is more robust to model misspecification.
The asymptotic properties and selection consistency of GIC are derived under the high-dimensional setting. Compared to AIC and BIC, the proposed GIC is better capable
than AIC in excluding noise eigenvalues, and is more sensitive than BIC in detecting signal eigenvalues. Moreover, we discuss an application of GIC to selecting the number of factors for factor analysis. Our numerical study reveals that GIC compares favorably to the methods based on (deterministic) parallel analysis.

\end{abstract}

\newpage

\section{Introduction}

Principal component analysis (PCA) is the most widely
used statistical procedure for dimension reduction. Let $X\in\real^p$ be a random vector
distributed from an arbitrary cumulative distribution function $G$ with mean $\mu$ and covariance $\Sigma$.
Consider the spectral decomposition $\Sigma = \Gamma\Lambda \Gamma^\top$, where $\Gamma=[\gamma_{1},\ldots,\gamma_{p}]\in \mathbb{R}^{p\times p}$
satisfies $\Gamma^\top \Gamma =I_p$, and $\Lambda\in \mathbb{R}^{p\times p}$ is a diagonal matrix
with positive diagonal entries arranged in descending order
\begin{equation}
\label{generalized_spiked_assumption}
\lambda_{1}> \lambda_{2}>\cdots > \lambda_{r_0}
\gg \lambda_{r_0+1} > \cdots > \lambda_{p}.
\end{equation}
Here we assume the existence of a sufficient gap between $\lambda_{r_0}$ and $\lambda_{r_0+1}$ so that $r_0$
can be treated as the {\it target rank} of $\Sigma$. With a rigorous definition of $r_0$, which will be given in
Section~\ref{sec.target_rank}, model~(\ref{generalized_spiked_assumption}) is called the {\it generalized spiked covariance model}. Then,
the leading eigenvectors $[\gamma_{1},\ldots,\gamma_{r_0}]$ are the target of PCA for dimension reduction.
In the sample level, let $\{X_i\}_{i=1}^n$ be a random sample, and let
$S_n=\frac{1}{n}\sum_{i=1}^{n}(X_i-\bar X)(X_i-\bar X)^\top$
be the sample covariance matrix, where $\bar X=\frac{1}{n}\sum_{i=1}^{n}X_i$. Consider the spectral decomposition
\begin{eqnarray}
S_n=\sum_{j=1}^p\hat\lambda_j\hat\gamma_j\hat\gamma_j^\top.\label{spectral_Sn}
\end{eqnarray}
If $r_0$ is known, then PCA suggests to map the data $\{X_i\}_{i=1}^n$ onto the space spanned by the leading eigenvectors $[\widehat\gamma_{1},\ldots,\widehat\gamma_{r_0}]$ for the subsequent analysis.
Since $r_0$ is usually unknown, an important issue of PCA is the determination of the target rank $r_0$.

There are many methods developed to determine $r_0$, and among them we mainly focus on the information-based criteria. Two commonly used information criteria for PCA rank
selection are the Akaike information criterion (AIC) and the Bayesian information criterion (BIC).
These methods adopt the number of free parameters as the penalty to prevent the selection of a saturated model.
The difference between the two criteria is that AIC uses a constant 2 to weight the model penalty,
while BIC uses $\ln n$. Statistical properties of AIC and BIC under the case of $n\to\infty$ and $p<\infty$ are studied in Fujikoshi and Sakurai (2016), wherein the authors show that AIC does not achieve selection consistency while BIC does. Later, the selection consistency of AIC and BIC under the case of $n\to\infty$ and $p/n\to c>0$ are established in Bai, Choi and Fujikoshi (2018) by using random matrix theory. Both the above-mentioned two works assume the validity of the {\it simple spiked covariance model}, i.e., model~(\ref{generalized_spiked_assumption}) with the constraint $\lambda_{r_0+1}=\cdots=\lambda_{p}$, for theoretical developments.

Due to its sparseness when $r_0 \ll p$,
the simple spiked covariance model provides a feasible alternative to its generalized counterpart~\eqref{generalized_spiked_assumption}.
However,
choosing $r_0$ via the model selection
criteria derived under the simple spiked covariance model (such as AIC and BIC) may no longer be suitable, provided some of the tail
eigenvalues in \eqref{generalized_spiked_assumption}
are distinct from each other.
This leads to the problem of PCA rank selection under model misspecification.
There is a rich literature on model selection under
misspecification; see Konishi and Kitagawa (1996),
Lv and Liu (2014), Hsu, Ing and Tong (2019) and the references therein.
Lv and Liu (2014) have introduced
the generalized BIC (GBIC) and
generalized AIC (GAIC) based on rigorous
asymptotic expressions of the Bayesian and KL divergence principles
in misspecified generalized linear models.
Hsu, Ing and Tong (2019) have
proposed the misspecification-resistant information criterion (MRIC) via an asymptotic expression
for the mean squared prediction error of a misspecified
time series model.
These criteria, however, focusing on regression-type models,
are not directly applicable to PCA rank selection.
Konishi and Kitagawa (1996), on the other hand,
have established an asymptotic expression
of the KL divergence principle in a general misspecification framework,
leading to their celebrated generalized information criterion (GIC).
Although the framework considered in Konishi and Kitagawa (1996) is quite general,
their derivation of GIC is reliant on some high-level assumptions,
which may fail to hold for PCA; see Section~\ref{sec.derivation_gic} for details.

In this paper, we derive the GIC-based bias correction, $b_r^\gic$,
to the KL divergence for the PCA rank selection problem, when a misspecified simple spiked covariance model is postulated.
While $b_r^\gic$ borrows the general idea of Konishi and Kitagawa (1996),
it is obtained under mild moment and distributional assumptions, thereby alleviating
the difficulty mentioned in the previous paragraph.
Moreover, we show that $b_r^\gic$ has an elegant expression in terms of the sizes of covariance eigenvalues.
This expression not only provides a deeper understanding of
the KL divergence principle in misspecified PCA models,
but it also leads to a
model selection criterion with a penalty term
more adaptive to various
spiked covariance structures than AIC and BIC, which simply depend on the number of free parameters. In particular, the proposed GIC for PCA rank selection is more
capable than AIC in excluding noise eigenvalues, and is more sensitive than BIC in detecting signal eigenvalues. Moreover, we also discuss an application of our GIC to selecting the number of factors for factor analysis (FA). Numerical study reveals that GIC compares favorably to the methods based on deterministic parallel analysis (DPA) proposed by Dobriban and Owen (2019).

The rest of this article is organized as follows.
Our GIC-based PCA rank selection method is derived in Section~2, and its asymptotic properties are investigated in Section~3. Numerical studies and a data example are provided in Section~4. Section~5 presents the application of GIC to selecting the number of factors in FA. This article ends with a discussion in Section~6. All proofs are placed in the Appendix.


\section{GIC for PCA Rank Selection}\label{sec:main}

\subsection{Review of AIC and BIC}\label{sec.aic_bic}

The AIC for PCA rank selection starts from fitting $\{X_i\}_{i=1}^n$ with the {\it rank-$r$ simple spiked covariance model} as the working model:
\begin{eqnarray}
\Sigma_r = \Gamma_r\Lambda_r\Gamma_r^\top +\sigma_r^2Q_r,\quad 0\le r\le p-1,
 \label{simple_spike}
\end{eqnarray}
where $\Gamma_r=[\gamma_{1},\ldots,\gamma_{r}]$, $\Lambda_r={\rm diag}(\lambda_{1},\ldots,\lambda_{r})$ with $\lambda_{1} > \cdots > \lambda_{r}>\sigma_r^2> 0$, and $Q_r=I_p-\Gamma_r\Gamma_r^\top$. The case of $r=0$ is understood as $\Sigma_0=\sigma_0^2I_p$.
The benefit of model~(\ref{simple_spike}) is that $r$ can be directly explained as the target rank of interest, and a certain
information criterion can be straightforwardly applied to determining a suitable value of $r$.
Assume the Gaussian distribution assumption $X\sim N(\mu,\Sigma_r)$, and let
\begin{eqnarray}\label{theta_r}
\theta_r=\left(\vec(\Gamma_r)^\top, \lambda_{1},\ldots,\lambda_{r}, \sigma_r^2,  \mu^\top \right)^\top.
\end{eqnarray}
The log-likelihood of $\theta_r$ (up to a constant term) is given by
\[
\ell_n(\theta_r)=\int\ell(x|\theta_r)dG_n(x) \quad{\rm with}\quad \ell(x|\theta_r)=-\frac12\ln\left|\Sigma_r\right|
 - \frac12\tr\left\{ (x-\mu)(x-\mu)^\top  \Sigma_r^{-1}\right\},
\]
where $G_n$ is the empirical distribution of $\{X_i\}_{i=1}^n$. Straightforward calculation shows that the MLE of $\theta_r$, i.e., $\hat\theta_r = \argmax _{\theta_r}\ell_n(\theta_r)$, consists of
\begin{eqnarray}\label{theta_r.MLE}
\widehat\Gamma_r=[\hat\gamma_{1},\ldots,\hat\gamma_{r}],\quad
\widehat\Lambda_r={\rm diag}(\hat\lambda_1,\ldots,\hat\lambda_r),\quad
\widehat\sigma_r^{2}=\frac{1}{p-r}\sum_{j>
r}\hat\lambda_{j},\quad \widehat\mu=\bar X,
\end{eqnarray}
where $\hat\gamma_j$'s and $\hat\lambda_j$'s are defined in (\ref{spectral_Sn}). Since $\tr(S_n\hat\Sigma_r^{-1})=p$,
\begin{equation}\label{Sigma_hat}
\ell_n(\widehat\theta_r)=-\frac{1}{2}\ln|\widehat\Sigma_r|-\frac{p}{2}\quad{\rm with}\quad\widehat\Sigma_r = \widehat\Gamma_r\widehat\Lambda_r \widehat\Gamma_r^{\top} +\widehat\sigma_r^{2}\widehat Q_r,
\end{equation}
where $\widehat Q_r = I_p-\widehat\Gamma_r
\widehat\Gamma_r^{\top}$. Then, AIC estimates $r_0$ by
\begin{eqnarray}
\widehat r_\aic = \argmax_{r\le q} \Big\{ \ell_n(\widehat\theta_r)
-\frac{1}{n}\, b_r\Big\}
= \argmin_{r\le q} \Big\{ \ln|\widehat\Sigma_r|
+\frac{2}{n}\, b_r\Big\}, \label{AIC}
\end{eqnarray}
where $q\le p-1$ is a pre-determined upper bound of the model rank, and
\begin{eqnarray}\label{b_r}
b_r = \left\{pr-\frac{r(r+1)}{2}\right\} +r+1 + p\label{number_para}
\end{eqnarray}
is the number of free parameters in $\theta_r$. BIC estimates $r_0$ in a similar manner with AIC, except that BIC
uses $\ln n$ as the weight for $b_r/n$:
\begin{eqnarray}
\widehat r_\bic = \argmin_{r\le q} \Big\{ \ln|\widehat\Sigma_r|
+\frac{\ln n}{n}\, b_r\Big\}. \label{BIC}
\end{eqnarray}
Statistical properties of $\widehat r_\aic$ and $\widehat r_\bic$ have been studied in Fujikoshi and Sakurai (2016) for finite $p$ case, and in Bai, Choi and Fujikoshi (2018) for diverging $p$ case, both assume that the true model is the simple spiked covariance model. We remind the reader that the true model considered in this paper is the generalized spiked covariance
model (\ref{generalized_spiked_assumption}), while the rank-$r$ simple spiked covariance model~(\ref{simple_spike}) is only a working model for rank selection.

Let $\theta_r^*$ be the population version of MLE from fitting the rank-$r$ simple spiked covariance
model~(\ref{simple_spike}). By~(\ref{theta_r.MLE}), we have that $\theta_r^*$ consists of
\begin{eqnarray}
\Gamma_r^*=[\gamma_{1},\ldots,\gamma_{r}],\quad
\Lambda_r^*={\rm diag}(\lambda_{1},\ldots,\lambda_{r}),\quad
\sigma_r^{*2}=\frac{1}{p-r}\sum_{j > r}\lambda_{j},\quad \mu^*=\mu,
\label{mle.theta_r.optimal}
\end{eqnarray}
and the rank-$r$ simple spiked covariance matrix induced by $\theta_r^*$ is given by
\begin{eqnarray*}
\Sigma_r^* = \Gamma_r^*\Lambda_r^* \Gamma_r^{*\top} +\sigma_r^{*2}Q_r^*,
\end{eqnarray*}
where $Q_r^*=I_p-\Gamma_r^*
\Gamma_r^{*\top}$. The notation $\theta_r^*$ and $\Sigma_r^*$ will be used in the subsequent discussion.

\begin{rmk}
Note that $\widehat\Sigma_r$ in (\ref{Sigma_hat}) also minimizes the log-determinant
divergence between $S_n$ and $\Sigma_r$: $\D( S_n  ,\Sigma_r) = -\frac12\ln\left| S_n  \Sigma_r^{-1}\right|
 + \frac12\tr\left( S_n  \Sigma_r^{-1}\right)-\frac{p}2$. Thus, if the Gaussian assumption is not valid, $\widehat \Sigma_r$ can still be explained as the minimum log-determinant divergence estimator.
\end{rmk}

\subsection{GIC for model complexity measure}\label{sec.derivation_gic}

AIC, penalizing model complexity based on the number of free parameters $b_r$, is obtained under
the crucial assumption that model~(\ref{simple_spike}) is the true data generating distribution. However, this simple spiked model
assumption can be easily violated in practical applications of high dimensional data.
When \eqref{simple_spike} is used as a working model but
the true model is \eqref{generalized_spiked_assumption},
the GIC of Konishi and Kitagawa (1996),
taking model misspecification into account,
appears to be a more appealing alternative.
To derive GIC in this case, one must establish an asymptotic expression for the bias incurred by using the within-sample
prediction error to approximate the likelihood-based prediction error,
\begin{eqnarray}
E\left\{\int\ell(x|\widehat\theta_r)dG_n(x) - \int\ell(x|\widehat\theta_r)dG(x) \right\},\label{gic_original}
\end{eqnarray}
which is suggested by the KL divergence principle.
Here, the expectation is taken with respect to
$X_1,\ldots,X_n\stackrel{\rm i.i.d.}{\sim}G$. Note that the true distribution $G$ need not be Gaussian.



Theorem 2.1 in Konishi and Kitagawa (1996) suggests that (\ref{gic_original}) can be
expressed as $n^{-1}b_r^\gic+o(n^{-1})$, where
\begin{eqnarray}\label{def:b_gic}
b_r^\gic = E\left[\left\{\frac{\partial\ell(X|\theta_r)}{\partial\theta_r}|_{\theta_r=\theta_r^*}\right\}^\top\IF_{T_r}(X;G)\right]
\end{eqnarray}
with the expectation being taken with respect to $X\sim G$. Here $\IF_{T_r}(x;G)$ is the influence function of the MLE functional $T_r$ from fitting the rank-$r$
simple spiked model~(\ref{simple_spike}), i.e., $\theta_r^*=T_r(G)$ and $\widehat\theta_r=T_r(G_n)$. The derivation
of Konishi and Kitagawa (1996), however, is based on the functional Taylor series expansion of $T_r$
around $G$ under the assumption that $T_r$ is second-order compact differentiable at $G$. The second-order compact
differentiability is difficult to check and can be violated for eigenvalues and eigenvectors as functionals of $G$. It is
known that even the sample mean is not compact differentiable (Beutner and Z\"ahle, 2010). Furthermore,
the proof of their Theorem~2.1 involves interchanging expectations
and probability limits, such as $\lim_{n\to \infty} E\left\{o_p(1)\right\}= 0$,
which is in general not true without imposing
some smoothness conditions on $G$. In Theorem~\ref{thm:bias} we give a rigorous derivation for prediction error bias correction for PCA rank selection and
show that the asymptotic expression given in Konishi and Kitagawa (1996) is still valid.

\begin{thm}\label{thm:bias}
Assume the following two conditions:
\begin{itemize}
\item[(i)]
$X$ has finite $4q^\th$ moments for some $q>1$.
\item[(ii)]
The distribution of $X$ satisfies the Lipschitz condition: there exist $c_o>0$, $0<\nu\le 1$, $0<\delta\le 1$
such that $\sup_{v\in\real^p, \|v\|=1} P\left( |v^\top (X-\mu)| \le\omega\right) \le  c_o\, (2\omega)^\nu$ for all $0<\omega \le\delta/2$.
\end{itemize}
Then, $\lim_{n\to\infty}nE\left\{\int\ell(x|\widehat\theta_r)dG_n(x) - \int\ell(x|\widehat\theta_r)dG(x) \right\}
=\, b_r^\gic$ for any fixed $r$ and $p$.
\end{thm}

\noindent
Whereas (i) imposes a mild moment restriction, assumptions like (ii) have been
frequently used to derive information criteria in a rigorous manner.
For example, under correct specification of the model and an assumption slightly stronger than (ii),
Findley and Wei (2002) have presented the first mathematically complete derivation
of AIC for vector autoregressions. As shown in Section~4 of Findley and Wei (2002),
their stronger assumption is fulfilled by
a rich class of multivariate distributions, including
the Gaussian, $t$, and $\varepsilon$-contaminated Gaussian. Note that we assume in Theorem~\ref{thm:bias} the finiteness of $p$ for mathematical tractability.
We are not able to extend the result to the case of diverging $p$. Nevertheless, to the best of our knowledge
there is no existent study for GIC-based PCA rank selection even for the fixed $p$ case.
The finiteness of $p$, however, is merely to establish Theorem~\ref{thm:bias}.
In the rest of the article, this assumption is no longer required in the development of our method nor for the selection consistency theorem.

It is worth emphasizing that $b_r^\gic$ consists of two parts: the differential of
log-likelihood $\partial\ell(x|\theta_r)/\partial\theta_r$ under the working model $N(\mu,\Sigma_r)$,
and the influence function $\IF_{T_r}(x;G)$ of MLE under the true $G$ with the generalized spiked covariance
matrix satisfies~(\ref{generalized_spiked_assumption}). Intuitively, $b_r^\gic$ can be estimated by
$\frac{1}{n}\sum_{i=1}^n\{\frac{\partial\ell(X_i|\widehat\theta_r)}{\partial\theta_r}\}^\top\IF_{T_r}(X_i;G_n)$
using empirical data. However, a closer look at the calculation of $b_r^\gic$ reveals that its estimation under the PCA
rank selection problem involves the fourth moments of $X$, which can be unstable in practice. Fortunately, a neat
expression of $b_r^\gic$ that avoids calculating higher-order moments of $X$ can be derived under the working assumption of Gaussianity on $X$.

\begin{thm}\label{thm.GIC.penalty}
Assume $X\sim N(\mu,\Sigma)$, where $\Sigma$ satisfies the generalized spiked covariance model~(\ref{generalized_spiked_assumption}). Then, we have $b_{r}^\gic=b_{\Gamma_r}^\gic+b_{\Lambda_r}^\gic+b_{\sigma_r^2}^\gic+b_{\mu}^\gic$, $0\le r\le p-1$, where
\begin{eqnarray*}
b_{\Gamma_r}^\gic
&=&  {r \choose 2} + \sum_{j\le r}\sum_{\ell>r}
  \frac{\lambda_{\ell}(\lambda_{j}-\sigma_r^{*2})}{\sigma_r^{*2}(\lambda_{j}-\lambda_{\ell})}, \\
b_{\Lambda_r}^\gic
&=& r, \\
b_{\sigma_r^2}^\gic &=&  \frac{\frac{1}{p-r}\sum_{j>r}\lambda_{j}^2}{ \left( \frac{1}{p-r}\sum_{j>r}\lambda_{j}\right)^2},\\
b_{\mu}^\gic &=& p
\end{eqnarray*}
account for the model complexity corresponding to $\Gamma_r$, $\Lambda_r$, $\sigma_r^2$, and $\mu$. Here ${r \choose 2}=0$ if $r\le1$, and $b_{\Gamma_r}^\gic=b_{\Lambda_r}^\gic=0$ if $r=0$.
\end{thm}

\noindent One can see from Theorem~\ref{thm.GIC.penalty} that  $b^\gic_{r}$ depends on the dispersion of $\lambda_{j}$'s, while the
value of $b_r$ does not. More insights for terms making up $b_r^\gic$ are listed below:

\begin{itemize}
\item
For $b^\gic_{\Gamma_r}$, recall that $pr -\frac{r(r+1)}2$ is the number of free parameters in $\Gamma_{r}$. We have
\begin{eqnarray*}
b^\gic_{\Gamma_{r}} - \left\{pr -\frac{r(r+1)}2\right\}
&=& \sum_{j\le r}\xi_{j| r} ,
\end{eqnarray*}
where
\[ \xi_{j| r} = \sum_{\ell > r} \left\{\frac{\lambda_{\ell}(\lambda_{j}-\sigma_{r}^{*2})}
  {\sigma_{r}^{*2}(\lambda_{j}-\lambda_{\ell})} -1\right\}
   = \frac{\lambda_j}{\sigma_r^{*2}}\sum_{\ell > r}\frac{\lambda_\ell-\sigma_r^{*2}}{\lambda_j-\lambda_\ell},\quad j\le r.\]
For a given $r$,  $\xi_{j|r}$ can be explained as the difference of model complexity for the $j$-th eigenvector between
models without and with the requirement of correctly specifying the rank-$r$ simple spiked model~(\ref{simple_spike}). By the definition of $\sigma_r^{*2}$ in (\ref{mle.theta_r.optimal}), we have
\begin{eqnarray*}
\xi_{j| r}&\ge &\frac{\lambda_j}{\sigma_r^{*2}}\sum_{\ell > r}\frac{\lambda_\ell-\sigma_r^{*2}}{\lambda_j-\sigma_r^{*2}}=0.
\end{eqnarray*}
This indicates that counting the number of free parameters generally underestimates the model complexity of $\Gamma_r$,
especially when $\{\lambda_\ell:\ell>r\}$ deviates from their mean $\sigma_r^{*2}$.

\item
For $b^\gic_{\sigma_r^2}$, we have
\[b^\gic_{\sigma_r^2}= \frac{\frac{1}{p-r}\sum_{j>r}\lambda_{j}^2}{\left(\frac{1}{p-r}\sum_{j>r}\lambda_{j} \right)^2}\ge 1,\]
where the equality holds if and only if the elements of $\{\lambda_{j}:j>r\}$ are all equal. That is to say, except for the case
$\Sigma=\Sigma_{r_0}$, where the population covariance exactly matches the simple spiked working covariance, counting the number of free parameters
underestimates the model complexity for tail components. However, $b^\gic_{\sigma_r^2}$ can be more adaptive to the eigenvalue dispersion in assessing
the complexity for tail components.

\item
For $b^\gic_{\Lambda_r}$ and $b^\gic_{\mu}$, GIC counts the number of free parameters in $\Lambda_r$ and $\mu$. It indicates that the
misspecification of a generalized spiked model by a simple spiked model has no impact on the complexity of leading $r$ eigenvalues and mean vector.
\end{itemize}
From the observations above, we have
\begin{equation}\label{GIC_AIC_r}
b_{r}^\gic \ge b_{r}\quad\forall~ (r,\theta_r).
\end{equation}
A trivial case for the equality to hold is the case of full model $r=p-1$, in which situation we have $b^\gic_{p-1}=b_{p-1}=\frac{1}{2}p(p+1)+p$. For $r<p-1$, the equality in (\ref{GIC_AIC_r}) holds if and only if $\Sigma=\Sigma_{r_0}$ and $r\ge r_0$ (see Remark~\ref{rmk.br_gic_equal_br} for details). As the simple spiked model~(\ref{simple_spike}) is rarely true, (\ref{GIC_AIC_r}) implies that $b_r$ generally underestimates the model complexity, in which situation model misspecification can be an issue.

With the expression of $b_r^\gic$, our GIC-based rank estimator is proposed to be
\begin{eqnarray}
\widehat r_{\gic} &=&
\argmin_{r\le q}
\Big\{\ln|\widehat\Sigma_r|  +\frac{2}{n}\, \widehat b_{r}^\gic\Big\}, \label{GIC}
\end{eqnarray}
where
\begin{eqnarray}
\hat b_{r}^\gic=\left\{{r \choose 2} + \sum_{j\le r}\sum_{\ell>
r}\frac{\hat\lambda_{\ell}(\hat\lambda_{j}-\widehat\sigma_r^{2})}{\widehat\sigma_r^{2}(\hat\lambda_{j}-\hat\lambda_{\ell})}\right\} + r +
\frac{\frac{1}{p-r}\sum_{j>r}\hat\lambda_{j}^2}{(\frac{1}{p-r}\sum_{j>r}\hat\lambda_{j})^2} + p
\label{GIC.sample}
\end{eqnarray}
is the sample version of $b^\gic_{r}$ obtained by plugging in the sample eigenvalues, and $q$ is a pre-determined upper bound of the model rank.
Note that $n^{-1}(b^\gic_{\Lambda_r}+ b^\gic_{\sigma_r^2}) =o(1)$, while $n^{-1}b^\gic_{\Gamma_r} =O(1)$, and that $b^\gic_{\mu}$ is independent
of the model rank $r$. This indicates that $b^\gic_{\Gamma_r}$ plays a dominant role in applying $\widehat r_{\gic}$. The form of $b_{\Gamma_r}^\gic$ also indicates that GIC will prevent the selection of a model rank with nearly multiple eigenvalues, in which situation a large value of $\widehat b^\gic_{\Gamma_r}$ is induced due to the division by $(\hat\lambda_{j}-\hat\lambda_{\ell})$. This scenario, however, cannot be reflected by $b_r$ that simply counts the number of free parameters.

We close this section by noting that the Gaussian assumption in Theorem~\ref{thm.GIC.penalty} is merely used to get an explicit neat expression
for $b_r^\gic$. This assumption is not critical in applying $\widehat r_\gic$. In Section~\ref{sec.asymptotic}, we will rigorously investigate the
asymptotic properties of $\widehat b_r^\gic$ and $\widehat r_{\gic}$ under the high-dimensional setting and without requiring the
Gaussian assumption.

\begin{rmk}\label{rmk.br_gic_equal_br}
The derivation of $b_r^\gic$ in Theorem~\ref{thm.GIC.penalty} is under the assumption of distinct eigenvalues~(\ref{generalized_spiked_assumption}).
For $\Sigma=\Sigma_{r_0}$, where the tail eigenvalues are equal, $b_r^\gic$ can still be well-defined by setting $0/0=1$ for these terms $\frac{\lambda_{j}-\sigma_r^{*2}}{\lambda_{j}-\lambda_{\ell}}$. With this treatment, it is straightforward to show that $b_r^\gic=b_r$ if and only if $\Sigma=\Sigma_{r_0}$ and $r\ge r_0$.
\end{rmk}

\section{Rank Selection Consistency}\label{sec.asymptotic}

\subsection{The target rank}\label{sec.target_rank}

The selection consistency is meaningful only when there is a precise definition for the target rank $r_0$ of $\Sigma$. The aim of this subsection is to provide a
rigorous definition for $r_0$ that is identifiable under model~(\ref{generalized_spiked_assumption}).

We start by reviewing some basic results of random matrix theory. Define the empirical spectral distribution (ESD) of $\Sigma$ to be
\begin{equation*}
F^{\Sigma}(t) = \frac{1}{p}\sum_{j=1}^{p}\mathcal{I}\{\lambda_j\le t\},
\end{equation*}
where $\mathcal{I}\{\cdot\}$ is the indicator function. We remind the reader that the Gaussian assumption for $X$ is no longer required in the rest of discussion. Instead, we assume the following conditions (C1)--(C4), which are commonly used in random matrix
theory (Bai and Yao, 2012).
\begin{itemize}
\item[(C1)]
For each $i$, $X_i$ takes the form $X_i=\Sigma^{1/2}Z_i$, where $Z_i\in \mathbb{R}^p$ consists of i.i.d.~random variables with mean 0, variance 1,
and finite fourth moment.

\item[(C2)]
$p/n\to c\in(0,\infty)$.

\item[(C3)]
$F^\Sigma$ converges weakly to $H$ as $p\to \infty$, where $H$ has bounded support.

\item[(C4)]
The sequence of spectral norms of $\Sigma$ is bounded.
\end{itemize}
\noindent The distribution $H$ is called the limiting spectral distribution (LSD) of $\Sigma$.
The LSD of the sample covariance matrix $S_n$ exists and is determined by $(c,H)$.

\begin{thm}[Yao, Zheng and Bai, 2015]\label{lem.limit.fcH}
Given $(c,H)$. Then, $F^{S_n}$ converges to the \textit{generalized Mar\v{c}enko-Pastur}  distribution $F_{c,H}$ as $n\to\infty$. The Stieltjes
transform of $F_{c,H}$, $s_{c,H}(z)=\int\frac{1}{t-z}dF_{c,H}(t)$, $z\in \mathbb{C}^+$, is implicitly determined via $s_{c,H}(z)=\int\frac{dH(t)}{t\{1-c-cz\, s_{c,H}(z)\}-z}$.
Moreover, $F_{c,H}$ and $H$ share the same mean, i.e., $\int tdF_{c,H}(t)=\int tdH(t)= \mu_H$.
\end{thm}

A major difficulty of random matrix theory is that $F_{c,H}$ has no explicit form for an arbitrary $H$. One known result for $F_{c,H}$ is under the
{\it simple spiked covariance model} assumption:
\begin{itemize}
\item[(C5)]
$H(t)=\I\{t\ge \sigma^2\}$ for some $\sigma^2>0$.\label{simple_spiked_assumption}
\end{itemize}

\begin{thm}[Mar\v{c}enko-Pastur Law] \label{lem.LSD.spikded_model}
Assume further the simple spiked covariance model (C5) and given $c$. Then, the probability density function of $F_{c,H}$ is given by
\begin{eqnarray*}
f_{c,\sigma^2}(t)=\frac{1}{2\pi c\sigma^2t} \sqrt{(t-a)(b-t)}\cdot\I\{a\le t\le b\}
\end{eqnarray*}
with an additional point mass at $t=0$ having probability $1-1/c$ if $c>1$, where $a=\sigma^2(1-\sqrt{c})^2$ and $b=\sigma^2(1+\sqrt{c})^2$. The mean
under $F_{c,H}$ is $\mu_H=\sigma^2$.
\end{thm}

An eigenvalue $\lambda_{j}$ is called a \textit{generalized spike eigenvalue} if $\lambda_{j}$ does not belong to the support of $H$, denoted by
$\mathcal{S}(H)$. For the PCA rank selection problem, we assume that the generalized spike eigenvalues are larger
than $\sup \mathcal{S}(H)$, so that these spike eigenvalues can be of the major interest. To study the asymptotic behavior of $\hat{\lambda}_j$, define
\begin{equation*}
\psi(\lambda)= \lambda\left\{1+c\int\frac{t}{\lambda-t}dH(t)\right\}\quad {\rm and}\quad \psi'(\lambda)=\frac{\partial\psi(\lambda)}{\partial\lambda},
\quad \lambda \notin \mathcal{S}(H).
\end{equation*}
Note that $\psi(\cdot)$ depends on $(c,H)$. The following results are modified from Bai, Chen and Yao (2010), Bai and Yao (2012), and Yao, Zheng
and Bai (2015).

\begin{thm}\label{lem.limit}
Let $\lambda_j$ be a generalized spike eigenvalue with $\psi_j=\psi(\lambda_j)$, and let $b=\sup\mathcal{S}(F_{c,H})$.
\begin{enumerate}
\item[(i)]
$\psi_j \notin \mathcal{S}(F_{c,H})$ (i.e., $\psi_j>b$) if and only if $\psi'(\lambda_j)>0$.

\item[(ii)]
If $\psi'(\lambda_{j})>0$, then $\hat\lambda_j\stackrel{a.s.}{\to}\psi_j$.

\item[(iii)]
If $\psi'(\lambda_{j})\le 0$ and $j=o(p)$, then $\hat\lambda_j\stackrel{a.s.}{\to} b$.

\item[(iv)]
For any $k=o(p)$, $\frac{1}{p-k}\sum_{\ell>k}\hat\lambda_\ell\stackrel{a.s.}{\to} \mu_H$.
\end{enumerate}
\end{thm}

\noindent These results show that a generalized spike eigenvalue $\lambda_{j}$ is not guaranteed to be identifiable in general. The identifiability of the $j$th spike further requires $\psi'(\lambda_{j})>0$ so that the corresponding limiting value $\psi_j$ can be separated
from $\mathcal{S}(F_{c,H})$. Thus, a generalized spike eigenvalue $\lambda_{j}$ is called a \textit{distant spike} if $\psi'(\lambda_{j})>0$;
otherwise, it is called a \textit{close spike}. This definition also implies that for distant spikes
$\lambda_j>\lambda_k$, one has $\psi_j>\psi_k>b$.

We are now in a position to define the {\it target rank} $r_0$ of a PCA model. Under the generalized spiked covariance
model~(\ref{generalized_spiked_assumption}), we know from Theorem~\ref{lem.limit} that only distant spike eigenvalues can be separated
from $\mathcal{S}(F_{c,H})$. This motivates the following definition of $r_0$.

\begin{defn}\label{r0}
Given $(c,H)$, define $r_0=\sum_{j=1}^{p}\mathcal{I}\{\psi'(\lambda_{j})>0\}$ to be the number of distant spikes. We call $r_0$ the {\it target rank} of $\Sigma$ for PCA rank selection calibrated by $(c,H)$.
\end{defn}

\noindent The definition of $r_0$ implies that $\psi_{r_0}>b$; hence, the rank-$r_0$ PCA subspace is identifiable under $(c,H)$. Consequently,
for any given $(c,H)$, $r_0$ is the well-defined target rank of $\Sigma$, under which we can study the selection consistency of $\widehat r_\gic$.

\subsection{Selection consistency of GIC}\label{sec.consistency_GIC}

The asymptotic properties of $\widehat r_\gic$ heavily depend on the asymptotic increment of the GIC penalty when one increases
the model rank from $(j-1)$ to $j$:
\begin{eqnarray*}
\kappa_j = \lim_{n\to\infty}~ \frac{1}{n}\left(\hat
b_{j}^\gic-\hat b_{j-1}^\gic\right),\quad 1\le j \le p-1.
\end{eqnarray*}
Define the function
\begin{equation}\label{kappa}
\kappa(u) = c (u-1) E\left\{\frac{(T/\mu_H)}{u-(T/\mu_H)}\right\},\quad u\in [\,b/\mu_H,\infty),
\end{equation}
where the expectation is taken with respect to $T\sim F_{c,H}$. The function $\kappa(u)$ is well defined for $u>b/\mu_H$. As to the boundary point $b/\mu_H$, it is defined by $\kappa(b/\mu_H)=\lim_{u\to b/\mu_H} \kappa(u)$ and is assumed bounded. We have the following results for the asymptotic increment $\kappa_j$.

\begin{thm}\label{thm.Kj.limit}
Assume conditions (C1)--(C4).
\begin{enumerate}
\item[(i)]
$\kappa(u)$ is a strictly decreasing function. Moreover, $\kappa(u)\ge c$ and $\lim_{u\to\infty}\kappa(u)=c$.

\item[(ii)]
For $j=o(p)$, we have
\begin{eqnarray}
\kappa_j = \Bigg\{\begin{array}{ll}
              \kappa(\psi_j/\mu_H),& j\le r_0, \\
              \kappa(b/\mu_H),& j> r_0.
            \end{array}\label{Kj}
\end{eqnarray}
Note that the limiting value $\kappa(b/\mu_H)$ is independent of $j$.
\end{enumerate}
\end{thm}

\noindent The implications of Theorem~\ref{thm.Kj.limit} are threefold. First, we have
\begin{eqnarray}
\kappa_j\ge c\quad {\rm for}~j=o(p).  \label{Kj_c}
\end{eqnarray}
For the rank-$0$ PCA model, $\lim_{n\to\infty}\frac{1}{n}b_0^\gic=\lim_{n\to\infty}\frac{1}{n}b_0=c$. Since the asymptotic
increment of $\frac{1}{n}b_r$ is $c$ (Bai, Choi and Fujikoshi, 2018), (\ref{Kj_c}) implies that
\begin{eqnarray}
\lim_{n\to\infty}P\left(\widehat r_\gic \le \widehat r_\aic \right)=1.\label{gic_aic}
\end{eqnarray}
That is, GIC tends to select a smaller model than AIC. Second, $\lim_{u\to\infty}\kappa(u)=c$ implies that for the model rank with a large eigenvalue, GIC tends to use the same
penalty $c$ as AIC does. Third, $\kappa_j$ is an increasing function of $j$ that attains the maximum value $\kappa(b/\mu_H)$ when $j>r_0$. That is, GIC places a higher penalty on the model rank for a smaller eigenvalue, and places the maximum penalty $\kappa(b/\mu_H)$ when the eigenvalue is smaller than the minimum distant spike $\lambda_{r_0}$. This is reasonable since an
eigenvector associated with a smaller eigenvalue is more difficult to estimate, and one must pay a higher cost in its estimation.
Thus, when there exists a sufficiently large gap between $\lambda_{r_0}$ and $\mathcal{S}(H)$ in
the sense that $\psi_{r_0}$ and $b$ are well separated, $\kappa(b/\mu_H)$ for $j>r_0$ is expected to be
sufficiently larger than $\kappa(\psi_{r_0}/\mu_H)$. That is, one will undertake a large penalty when the model size goes beyond $r_0$,
and this property will drive GIC to select the target rank $r_0$. Below we state the selection consistency of GIC.

\begin{thm}\label{thm.GIC.consistency}
Assume conditions (C1)--(C4) and $q=o(p)$. Then, $\widehat r_\gic$ achieves selection consistency if and only if the following conditions are satisfied:
\begin{enumerate}
\item[(G1)] $L_\gic(\psi_{r_0}/\mu_H)<0$,
\item[(G2)] $L_\gic(b/\mu_H)>0$,
\end{enumerate}
where $L_\gic(u) = \ln u-(u-1)+2\kappa(u)$, and $\kappa(u)$ is defined in (\ref{kappa}).
\end{thm}

\noindent
Note that ``$\ln u-(u-1)$'' in $L_\gic(u)$ corresponds to a decrement in negative log-likelihood and ``$\kappa(u)$'' corresponds to
an increment of model penalty, when the rank of model covariance increases. Following the terminology of Bai, Choi and Fujikoshi (2018),
we call (G1)--(G2) the {\it gap conditions of GIC}. Condition (G1) gives the minimum size for the distant spike eigenvalue $\lambda_{r_0}$,
while condition (G2) quantifies the maximum size for the upper bound $b$ of $\mathcal{S}(F_{c,H})$. Recall from Theorem~\ref{thm.Kj.limit}
that $\kappa(\cdot)$ achieves the maximum value at $\kappa(b/\mu_H)$, which implies that (G2) is usually not as critical as (G1) to the selection consistency of GIC.


The selection consistency of AIC and BIC under the simple spiked covariance model with diverging $p$ has been established in Bai, Choi and Fujikoshi (2018).
Following the proof of Theorem~\ref{thm.GIC.consistency}
with $2\kappa(\cdot)$ therein replaced by $2c$ and $(\ln n)c$,
we can extend the results of Bai, Choi and Fujikoshi (2018) to the generalized
spiked covariance model~(\ref{generalized_spiked_assumption}).

\begin{thm}\label{corr.consistency}
Assume conditions (C1)--(C4) and $q=o(p)$.
\begin{enumerate}
\item[(i)]
$\widehat r_\aic$ achieves selection consistency if and only if the following conditions are satisfied:
\begin{enumerate}
\item[(A1)] $L_\aic(\psi_{r_0}/\mu_H)<0$,
\item[(A2)] $L_\aic(b/\mu_H)>0$,
\end{enumerate}
where $L_\aic(u)=\ln u-(u-1)+2c$.

\item[(ii)]
$\widehat r_\bic$ achieves selection consistency if and only if the following conditions are satisfied:
\begin{enumerate}
\item[(B1)] $\lim_{n\to\infty} L_\bic(\psi_{r_0}/\mu_H) < 0$,
\item[(B2)] $\lim_{n\to\infty} L_\bic(b/\mu_H) > 0$,
\end{enumerate}
where $L_\bic(u)=\ln u-(u-1)+(\ln n)c$.
\end{enumerate}
\end{thm}

Some further observations for these selection criteria are listed below:

\begin{itemize}
\item
For AIC, $\kappa_j\ge c$ from (\ref{Kj_c}) implies that the gap condition (G1) of GIC is stronger than
(A1) of AIC (denoted by ``$({\rm G1})\succ({\rm A1})$'' in the rest of the discussion).
This is reasonable since the penalty of GIC involves the estimation of eigenvalues,
and a larger $\psi_{r_0}/\mu_H$ is required to make the estimation of $b_r^{\rm GIC}$ stable, whereas the AIC penalty involves only counting the number of free parameters without the need to estimate them.
The benefit of adopting eigenvalues into the model penalty is that it makes GIC more robust to the form of $H$ than AIC. In particular, the validity of (G2) is less affected by a large value of $b$ due to the factor $2\kappa(b/\mu_H)$ (recall that $\kappa(\cdot)$ is a decreasing function and $\kappa(\cdot)\ge c$), while (A2) adopts a constant $2c$ for all $b$.
In other words, we have ``$({\rm A2})\succ ({\rm G2})$''. The advantage of GIC comes from its robustness in that $b_r^\gic$ does not require
the rank-$r$ working model~(\ref{simple_spike}) is the true model. GIC can thus be more adaptive to various structures of $\Sigma$ than AIC,
especially when the working model~(\ref{simple_spike}) is heavily violated.

\item
For BIC, it can be seen that (B2) is automatically satisfied for sufficiently large $n$, which implies ``$({\rm G2})\succ ({\rm B2})$''.
Although this indicates the robustness of BIC to the size of $b$, the cost is that the validity of (B1) requires a diverging $\psi_{r_0}/\mu_H$,
i.e., ``$({\rm B1})\succ ({\rm G1})$''. This phenomenon is also found in Bai, Choi and Fujikoshi (2018),
wherein the authors develop the selection consistency of BIC when $\lim_{n\to\infty}\{(\psi_{r_0}/\mu_H)/\ln n\}=\infty$. Briefly,
the validity of BIC requires a large sample size $n$ (such that (B2) is satisfied) and an extremely large value of $\psi_{r_0}/\mu_H$
(such that (B1) is satisfied), which can cause BIC to fail to identify $r_0$ when $n$ and $\psi_{r_0}/\mu_H$ have only moderate sizes. Unlike BIC, which uses $(\ln n)c$ in (B1)--(B2), GIC uses in (G1)--(G2) the function $2\kappa(\cdot)$, which is adaptive to $b$ and does not depend on $n$. As a result, GIC can achieve selection consistency even with finite $\psi_{r_0}/\mu_H$.
\end{itemize}
In summary, GIC can be treated as an intermediate method between AIC and BIC, by observing that ``$({\rm B1})\succ ({\rm G1})\succ ({\rm A1})$'' and ``$({\rm A2})\succ ({\rm G2})\succ ({\rm B2})$'' (see also Remark~\ref{rmk.bic_gic_aic} for more explanation). On one side, AIC is less affected by the small size of signal eigenvalues, but heavily depends on the correctness of the simple spiked model assumption, which can fail to apply for general $H$. On the other side, BIC is less affected by the form of $H$ due to the diverging factor $\ln n$, but it can fail to apply when the signal eigenvalues are not large enough. As for GIC penalty, it is constructed under the generalized spiked covariance model~(\ref{generalized_spiked_assumption}) that uses the adaptive $\kappa(b/\mu_H)$ to control the effect of $H$, in which situation GIC achieves selection consistency with moderate signal eigenvalues. We will further examine these issues via numerical studies in Section~\ref {sec.sim}.

\begin{rmk}\label{rmk.bic_gic_aic}
BIC adopts a diverging weight $\ln n$ for $b_r/n$, which implies that $(\ln n)c > 2\kappa(\cdot) \ge 2c$ for $n$ large enough. This extends (\ref{gic_aic}) to $\lim_{n\to\infty}P\left(\widehat r_\bic\le \widehat r_\gic \le \widehat r_\aic \right)=1$. That is, GIC tends to select a model of size between BIC and AIC.
\end{rmk}

\subsection{The case of simple spiked covariance model}

To get a better understanding of GIC behavior, this subsection concerns the special case of the simple spiked model given by (C5). Under (C5), $\psi(\lambda)=\lambda(1+\frac{c\,\sigma^2}{\lambda-\sigma^2})$, and hence, $\psi'(\lambda_j)>0$ if $\lambda_j>\sigma^2(1+\sqrt{c})$. Thus, $\lambda_{j}>\sigma^2(1+\sqrt{c})$ for $j\le r_0$. Moreover, the LSD of $S_n$ has an explicit form in Theorem~\ref{lem.LSD.spikded_model} with $b=\sigma^2(1+\sqrt{c})^2$. These properties further enable us to derive an explicit expression for $\kappa_j$ as stated below.

\begin{thm}\label{thm.Kj.limit.simple_spike}
Assume conditions (C1)--(C5). Then, for $j=o(p)$, we have
\begin{eqnarray*}
\kappa_j = \Bigg\{\begin{array}{ll}
            c+\frac{c^2\lambda_{j}/\sigma^2}{(\lambda_{j}/\sigma^2-1)^2},\quad j\le r_0,\\
            2c+c\sqrt{c},~~\quad\quad j > r_0.
            \end{array}\label{Kj_simple_spike}
\end{eqnarray*}
For $j\le r_0$, the limit of $\kappa_j$ goes to $2c+c\sqrt{c}$,  when $\lambda_{j}$ decreases to the critical value $\sigma^2(1+\sqrt{c})$.
\end{thm}

\begin{figure}[!ht]
\hspace{-2.2cm}
\includegraphics[width=8in,height=3.6in]{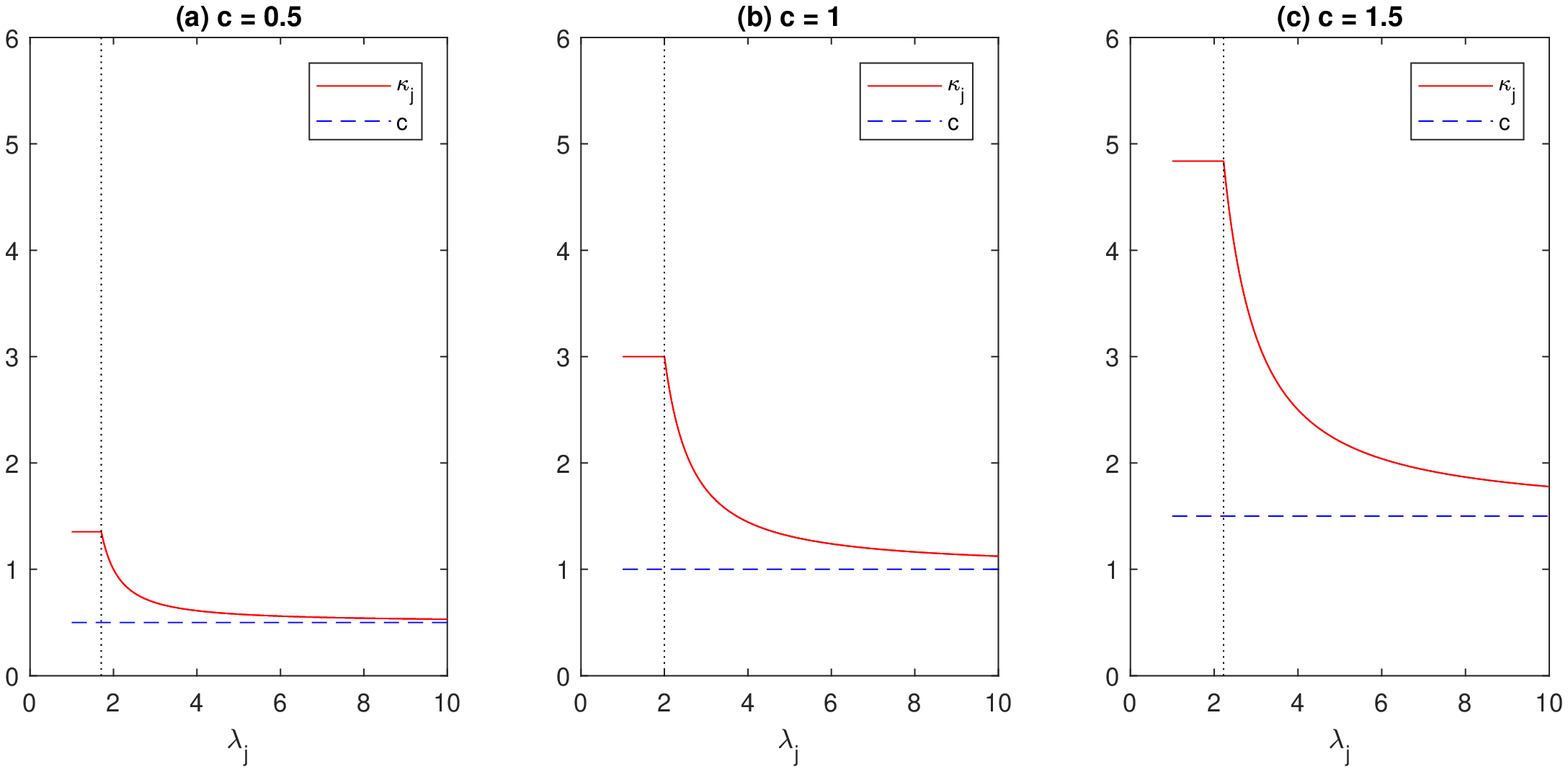}
\caption{The asymptotic increment $\kappa_j$ of $b_r^\gic$ as a function of $\lambda_j\in[1,\infty)$
under the simple spiked model (C5) with $\sigma^2=1$, where the vertical dotted line represents the lower bound $(1+\sqrt{c})$ for distant spiked eigenvalues. $\kappa_j$ achieves the maximum value $(2c+c\sqrt{c})$ for $\lambda_j\in[1,1+\sqrt{c}\,]$, and decreases to $c$ as $\lambda_j\to\infty$. The horizontal dashed line corresponds to the asymptotic increment of $b_r$ with a constant height $c$. (a) $c=0.5$; (b) $c=1$; (c) $c=1.5$.} \label{fig.kappa_j}
\end{figure}

\noindent Theorem~\ref{thm.Kj.limit.simple_spike} explicitly quantifies the dependence of $\kappa_j$ on the values of $(\lambda_{j}/\sigma^2,c)$ under the simple spiked model. Figure~\ref{fig.kappa_j} presents plots of $\kappa_j$ as a function of $\lambda_{j}\in[1,\infty)$ under (C5) with $\sigma^2=1$ and $c\in\{0.5,1,1.5\}$ . One can see that $\kappa_j$ achieves the maximum value $(2c+c\sqrt{c})$ for non-identifiable size of eigenvalues (i.e., $\lambda_{j}\in [1,1+\sqrt{c}\,]$), and decreases to the lower bound $c$ as $\lambda_{j}\to\infty$. This also supports our previous findings that
$b_r^\gic \ge b_r$, and that $b_r^\gic$ and $b_r$ tend to use the same penalty $c$ for spikes with large $\lambda_j$ values.

Under the simple spiked covariance model, Bai, Choi and Fujikoshi (2018) show that (A2) of AIC is automatically satisfied and, hence, AIC achieves selection consistency if and only if (A1) is satisfied. Since $\kappa(\cdot)\ge c$, (G2) of GIC is also satisfied under (C5). Thus, GIC achieves selection consistency if and only if (G1) is satisfied. This is stated in the following theorem. The proof is a direct consequence of Theorem~\ref{thm.GIC.consistency} and Theorem~\ref{thm.Kj.limit.simple_spike} and is omitted.

\begin{thm}\label{thm.GIC.consistency.simple_spike}
Assume conditions (C1)--(C5) and $q=o(p)$. Then, GIC achieves selection consistency if and only if
\begin{enumerate}
\item[(G1)$'$] $\ln(\psi_{r_0}/\sigma^2)-(\psi_{r_0}/\sigma^2-1)+2\left\{c+\frac{c^2\lambda_{r_0}/\sigma^2}{(\lambda_{r_0}/\sigma^2-1)^2}\right\} <0$,
\end{enumerate}
where $\psi_{r_0}=\lambda_{r_0} \Big(1+\frac{c\sigma^2}{\lambda_{r_0}-\sigma^2}\Big)$.
\end{thm}

\noindent Comparing (G1)$'$ with (A1), the required condition of GIC involves an extra term
\begin{equation}\label{penalty_diff}
\frac{2 c^2\lambda_{r_0}/\sigma^2}{(\lambda_{r_0}/\sigma^2-1)^2} .
\end{equation}
One message is that under the simple spiked covariance model, $c$ plays an important role in the performance of GIC. When $c<1$, the difference (\ref{penalty_diff}) is less critical; hence, both GIC and AIC achieve selection consistency for moderate size of $\lambda_{r_0}/\sigma^2$. When $c$ goes farther beyond 1, the difference (\ref{penalty_diff}) increases quadratically; hence, GIC requires a relatively large $\lambda_{r_0}/\sigma^2$ to ensure the selection consistency as compared with AIC. This phenomenon can also be seen in Figure~\ref{fig.kappa_j}, where the difference between $\kappa_j$ and $c$ grows nonlinearly as $c$ increases.
We should emphasize that the argument above holds for the simple spiked covariance model only. For general $H$, (A2) of AIC can be violated, while (G2) of GIC can be more easily satisfied because $\kappa(b/\mu_H)\ge c$. $F_{c,H}$ rarely has an explicit form for general $H$, which complicates the theoretical comparison of GIC with conventional methods. In Section~\ref{sec.sim}, we use simulation to compare the performance for different methods under various settings of $(c,H)$.

\section{Simulation Studies and a Data Analysis}\label{sec.sim}

\subsection{Simulation settings}

We consider three settings of $H$ for data generation:
\begin{itemize}
\item[(H1)] A simple spiked covariance model with $H(t)=\I\{t\ge 1\}$. It gives $\psi(\lambda)=\lambda\left(1+\frac{c}{\lambda-1}\right)$.

\item[(H2)] A generalized spiked covariance model with $H$ being a continuous uniform distribution on $[1-\theta, 1+\theta]$. It gives $\psi(\lambda)=\lambda\left[1+c\left\{\frac{\lambda}{2\theta}\ln\left(\frac{\lambda-1+\theta}{\lambda-1-\theta}\right)-1\right\}\right]$.

\item[(H3)] A generalized spiked covariance model with $H$ being a discrete uniform distribution on $\{1-\theta,1,1+\theta\}$. It gives $\psi(\lambda)=\lambda\left[1+\frac{c}{3}\left\{\frac{1-\theta}{\lambda-(1-\theta)}+\frac{1}{\lambda-1}+\frac{1+\theta}{\lambda-(1+\theta)}\right\}\right]$.
\end{itemize}
Note that $\mu_H=1$ under (H1)--(H3), and $\theta\in(0,1)$ in (H2)--(H3) is a quantity measuring the deviation from the simple spiked covariance model. For each $(c,H)$ combination, we consider two settings of distant spike eigenvalues:
\begin{itemize}
\item[(L1)]
$\lambda_{j}=\{1+
\frac{1}{10}(r_0-j+1)\}\lambda_{\gic}$, $j=1,\ldots,r_0$, where $\lambda_{\gic}=\lambda_{\gic}(c,H)$ is the unique solution for $L_\gic\{\psi(\cdot)\}=0$. It implies that condition (G1) of GIC is satisfied.

\item[(L2)]
$\lambda_{j}=\{1+\frac{1}{10}(r_0-j+1)\}\lambda_{\bic}$, $j=1,\ldots,r_0$, where $\lambda_{\bic}=\lambda_{\bic}(c,H)$ is the unique solution for $L_\bic\{\psi(\cdot)\}=0$. It implies that condition (B1) of BIC is satisfied.
\end{itemize}
The remaining eigenvalues $\{\lambda_{j}\}_{j=r_0+1}^p$ are generated from $H$. Similarly, let $\lambda_\aic=\lambda_\aic(c,H)$ be such that $L_\aic\{\psi(\lambda_{\aic})\}=0$.
Since $\kappa(\cdot)\ge c$, we have $\lambda_{\gic}\ge\lambda_{\aic}$. Hence, provided that $\ln n> 2$, the gap condition (A1) of AIC is also satisfied under (L1)--(L2). Finally, the data $\{X_i\}_{i=1}^n$ is generated from $N(0,\Sigma)$ with $\Sigma={\rm diag}(\lambda_{1},\ldots,\lambda_{p})$. We implement all methods with the candidate model size upper bound $q=20$. For each setting, the selection rate of a rank-$r$ model (over 200 replicates) is reported for $r\in\{0,1,2,\ldots,20\}$ under $r_0=5$, $n=500$, $\theta=0.8$, and $c\in\{0.5,1,1.5\}$.

\subsection{Simulation results}

\begin{figure}[!ht]
\hspace{-1.5cm}
\includegraphics[width=7.0in, height=7.2in]{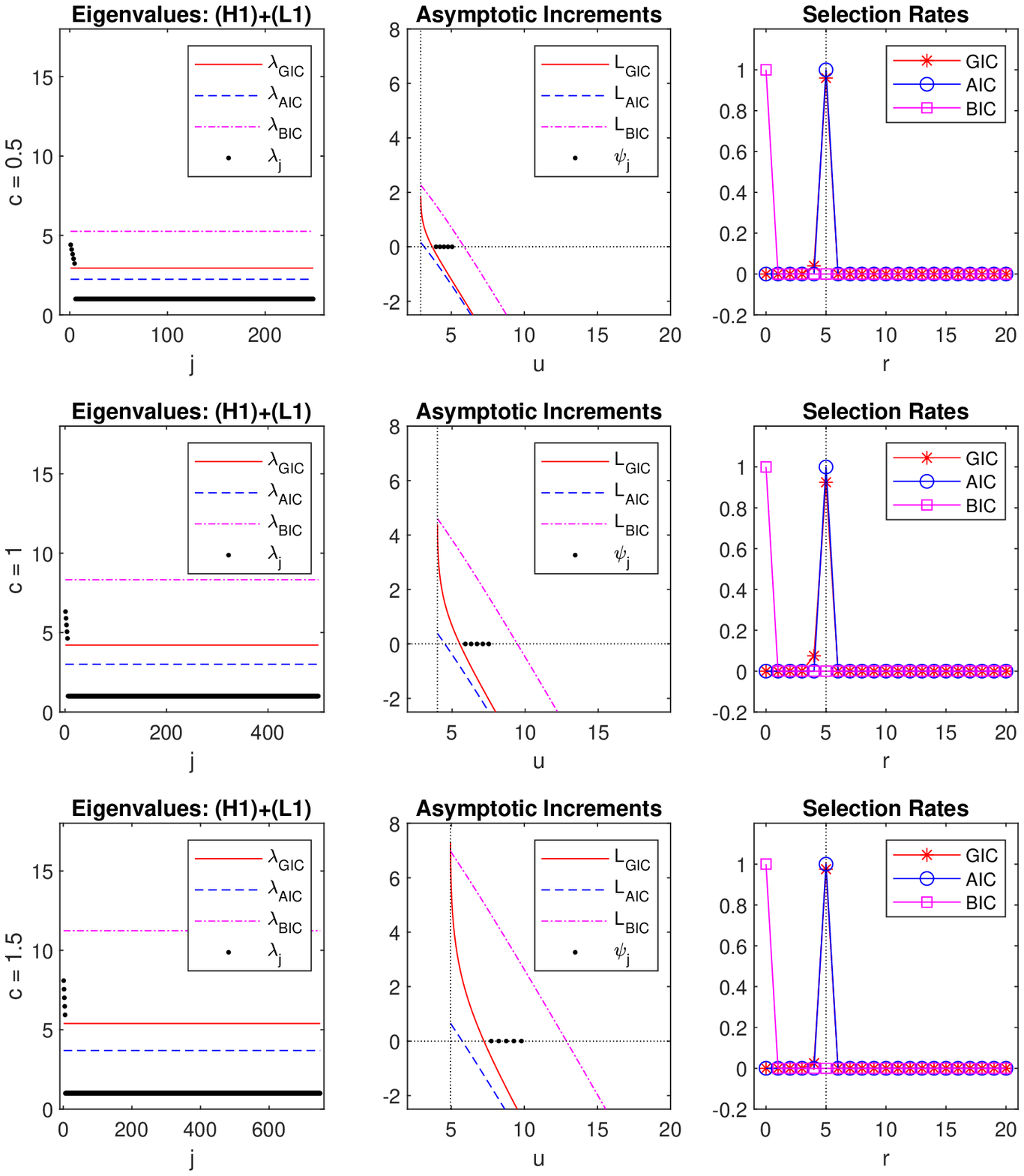}\\[-12ex]
\caption{Simulation results under (H1)+(L1) at different values of $c\in\{0.5,1,1.5\}$ (from top to bottom). The first column plots the true eigenvalues, where the horizontal lines represents $\lambda_\gic$ (the red solid line), $\lambda_\aic$ (the blue dashed line), and $\lambda_\bic$ (the magenta dash-dotted line). The second column plots the curves of $L_\gic$, $L_\aic$, and $L_\bic$, where the vertical dotted line represents the value of $b$, and the horizontal dotted line represents the value of 0. The values $\{\psi_j:j=1,\ldots,5\}$ of distant spikes are also reported. The third column reports the selection rate for each rank $r$, where the vertical dotted line indicates the true model size $r_0=5$.} \label{fig.sim_H1_L1}
\end{figure}

\begin{figure}[!ht]
\hspace{-1.5cm}
\includegraphics[width=7.0in, height=7.2in]{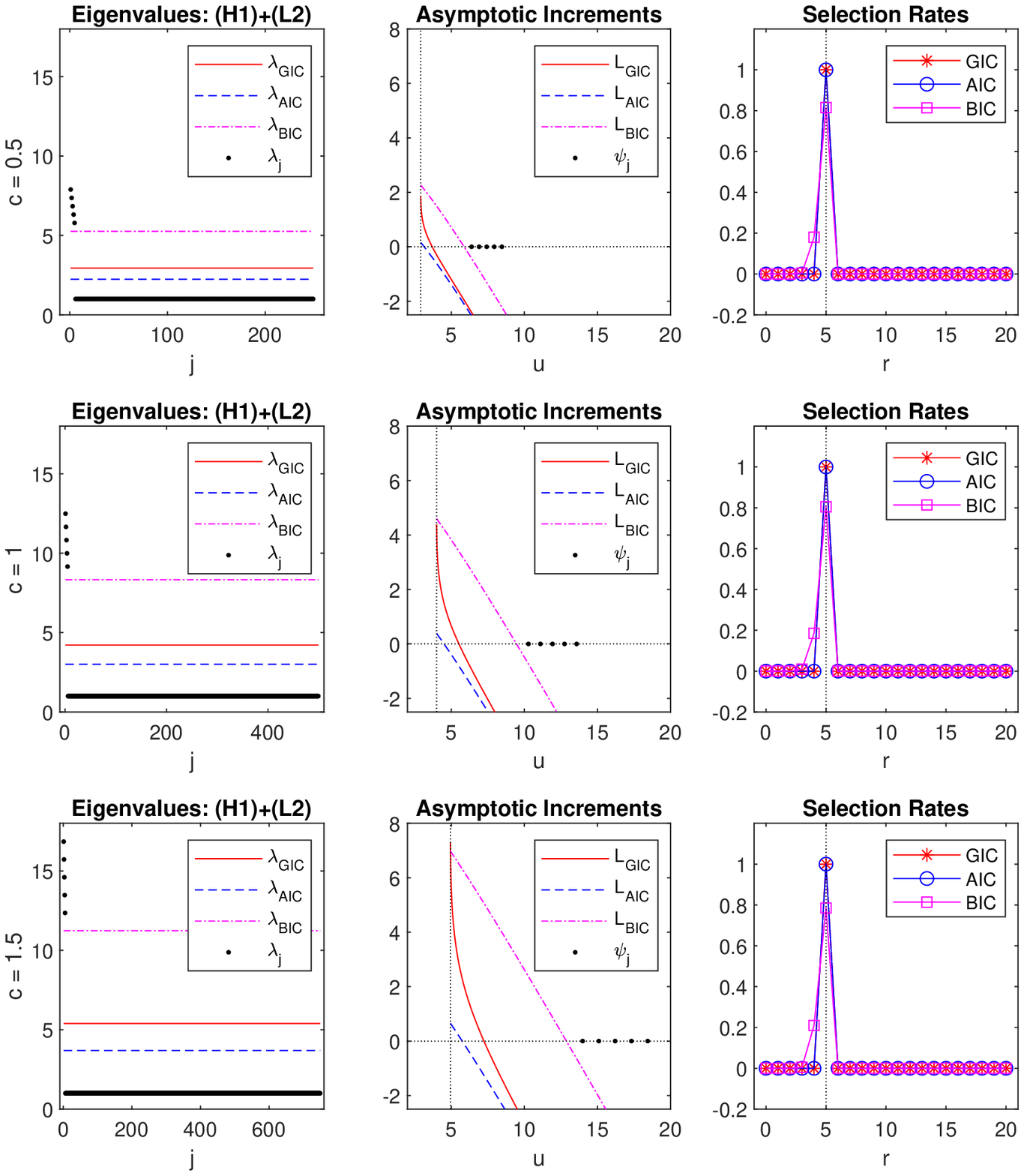}\\[-12ex]
\caption{Simulation results under (H1)+(L2) at different values of $c\in\{0.5,1,1.5\}$ (from top to bottom). The first column plots the true eigenvalues, where the horizontal lines represents $\lambda_\gic$ (the red solid line), $\lambda_\aic$ (the blue dashed line), and $\lambda_\bic$ (the magenta dash-dotted line). The second column plots the curves of $L_\gic$, $L_\aic$, and $L_\bic$, where the vertical dotted line represents the value of $b$, and the horizontal dotted line represents the value of 0. The values $\{\psi_j:j=1,\ldots,5\}$ of distant spikes are also reported. The third column reports the selection rate for each rank $r$, where the vertical dotted line indicates the true model size $r_0=5$.}
\label{fig.sim_H1_L2}
\end{figure}

Simulation results under (H1) are placed in Figures~\ref{fig.sim_H1_L1}--\ref{fig.sim_H1_L2}. The first column depicts the true eigenvalues $\{\lambda_{j}\}_{j=1}^p$ and the critical values of gap conditions $(\lambda_\gic, \lambda_\aic, \lambda_\bic)$.
One can see that $\lambda_\bic>\lambda_\gic>\lambda_\aic$ as expected, and $(\lambda_\gic,\lambda_\aic,\lambda_\bic)$ are all increasing functions of $c$, indicating that conditions (G1), (A1), and (B1) become stricter as $c$ increases. The second column of Figures~\ref{fig.sim_H1_L1}--\ref{fig.sim_H1_L2} depicts the curves $\{L_\gic(u):u\ge b\}$, $\{L_\aic(u):u\ge b\}$, and $\{L_\bic(u):u\ge b\}$, where the maximum value $b=(1+\sqrt{c})^2$ for $\mathcal{S}(F_{c,H})$ is represented as the vertical dotted line. Recall that $L_\gic\{\psi(\lambda_\gic)\}=0$, which means the  $L_\gic$ curve intersects the horizontal dotted line (the $x$-axis) at $u=\psi(\lambda_\gic)$, and (G1) of GIC is satisfied if $\psi_{r_0}$ lies on the right of $\psi(\lambda_\gic)$. Moreover, (G2) is satisfied if $L_\gic(b)>0$. Similar explanations apply to $L_\aic$ and $L_\bic$. Recall that Bai, Choi and Fujikoshi (2018) have shown that (A2) is satisfied under the simple spiked covariance model. This assertion can be observed in the simulation results, where $L_\gic(b)>L_\aic(b)>0$ for all $c\in \{0.5,1,1.5\}$. As a result, both GIC and AIC achieve selection consistency under (H1)+(L1) and (H1)+(L2). The selection rates are shown in the third column of Figures~\ref{fig.sim_H1_L1}--\ref{fig.sim_H1_L2}. It can be seen that both GIC and AIC consistently select rank-5 model. Note that condition (G1) is stronger than (A1) (i.e., $\lambda_\gic>\lambda_\aic$), and we can observe that AIC performs slightly better than GIC. As for BIC, it fails to identify the true rank $r_0$ under (H1)+(L1), but achieves selection consistency under (H1)+(L2). This is reasonable since condition (B1) is violated under (H1)+(L1) but is satisfied under (H1)+(L2) (the first column of Figures~\ref{fig.sim_H1_L1}--\ref{fig.sim_H1_L2}).

\begin{figure}[!ht]
\hspace{-1.5cm}
\includegraphics[width=7.0in, height=7.2in]{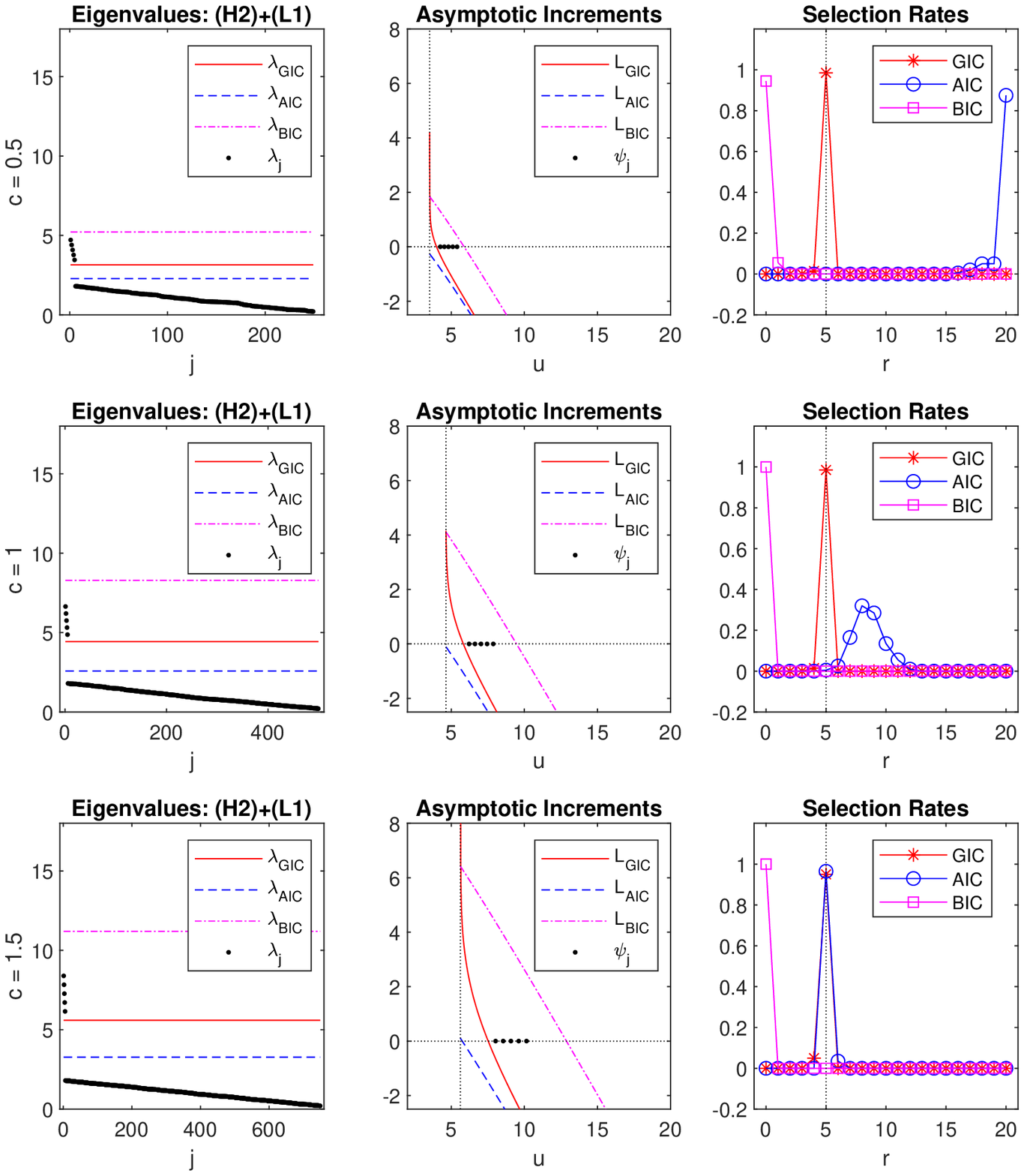}\\[-12ex]
\caption{Simulation results under (H2)+(L1) at different values of $c\in\{0.5,1,1.5\}$ (from top to bottom). The first column plots the true eigenvalues, where the horizontal lines represents $\lambda_\gic$ (the red solid line), $\lambda_\aic$ (the blue dashed line), and $\lambda_\bic$ (the magenta dash-dotted line). The second column plots the curves of $L_\gic$, $L_\aic$, and $L_\bic$, where the vertical dotted line represents the value of $b$, and the horizontal dotted line represents the value of 0. The values $\{\psi_j:j=1,\ldots,5\}$ of distant spikes are also reported. The third column reports the selection rate for each rank $r$, where the vertical dotted line indicates the true model size $r_0=5$.}
\label{fig.sim_H2_L1}
\end{figure}
\begin{figure}[!ht]
\hspace{-1.5cm}
\includegraphics[width=7.0in, height=7.2in]{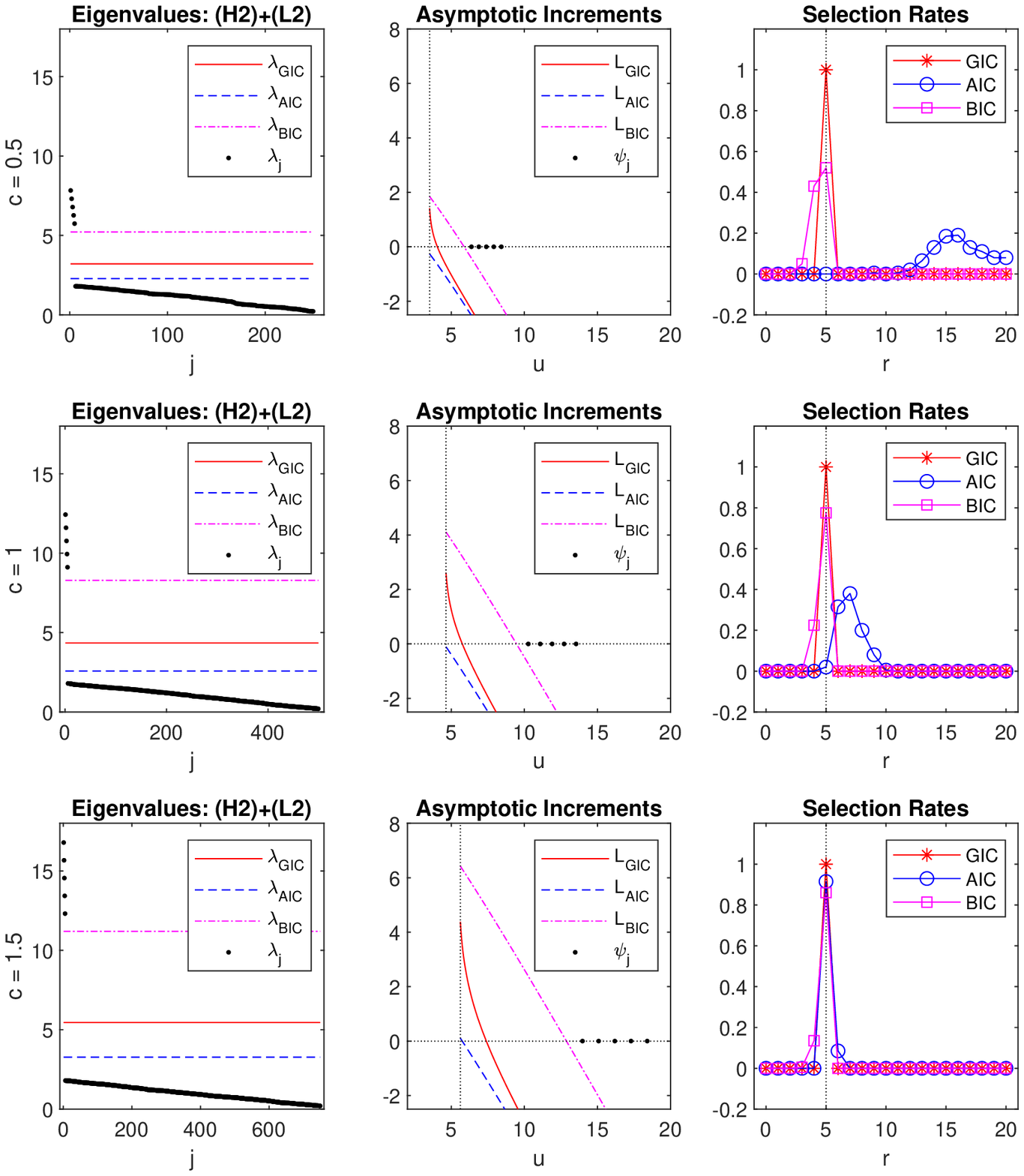}\\[-12ex]
\caption{Simulation results under (H2)+(L2) at different values of $c\in\{0.5,1,1.5\}$ (from top to bottom). The first column plots the true eigenvalues, where the horizontal lines represents $\lambda_\gic$ (the red solid line), $\lambda_\aic$ (the blue dashed line), and $\lambda_\bic$ (the magenta dash-dotted line). The second column plots the curves of $L_\gic$, $L_\aic$, and $L_\bic$, where the vertical dotted line represents the value of $b$, and the horizontal dotted line represents the value of 0. The values $\{\psi_j:j=1,\ldots,5\}$ of distant spikes are also reported. The third column reports the selection rate for each rank $r$, where the vertical dotted line indicates the true model size $r_0=5$.}
\label{fig.sim_H2_L2}
\end{figure}
\begin{figure}[!ht]
\hspace{-1.5cm}
\includegraphics[width=7.0in, height=7.2in]{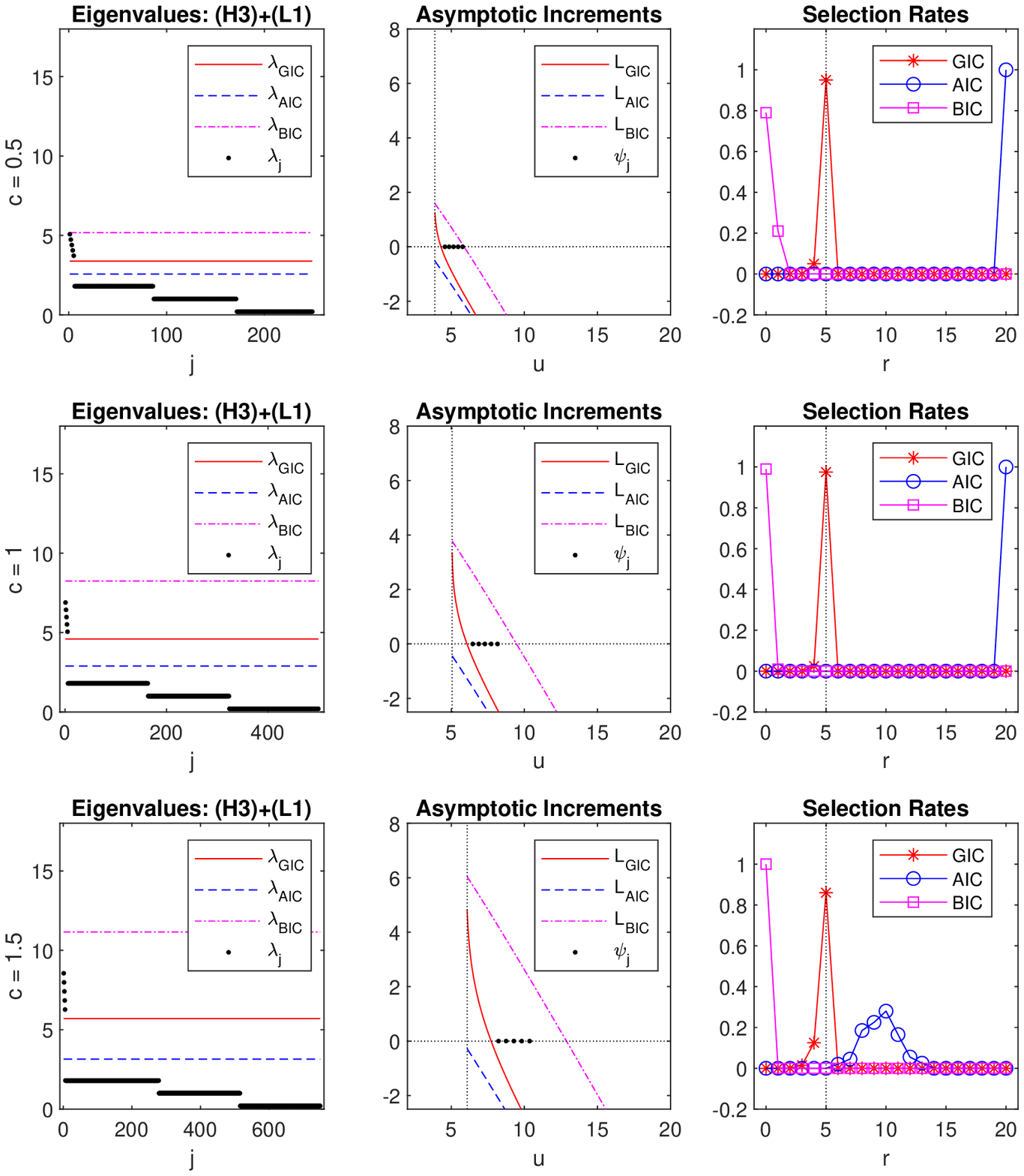}\\[-12ex]
\caption{Simulation results under (H3)+(L1) at different values of $c\in\{0.5,1,1.5\}$ (from top to bottom). The first column plots the true eigenvalues, where the horizontal lines represents $\lambda_\gic$ (the red solid line), $\lambda_\aic$ (the blue dashed line), and $\lambda_\bic$ (the magenta dash-dotted line). The second column plots the curves of $L_\gic$, $L_\aic$, and $L_\bic$, where the vertical dotted line represents the value of $b$, and the horizontal dotted line represents the value of 0. The values $\{\psi_j:j=1,\ldots,5\}$ of distant spikes are also reported. The third column reports the selection rate for each rank $r$, where the vertical dotted line indicates the true model size $r_0=5$.}
\label{fig.sim_H3_L1}
\end{figure}
\begin{figure}[!ht]
\hspace{-1.5cm}
\includegraphics[width=7.0in, height=7.2in]{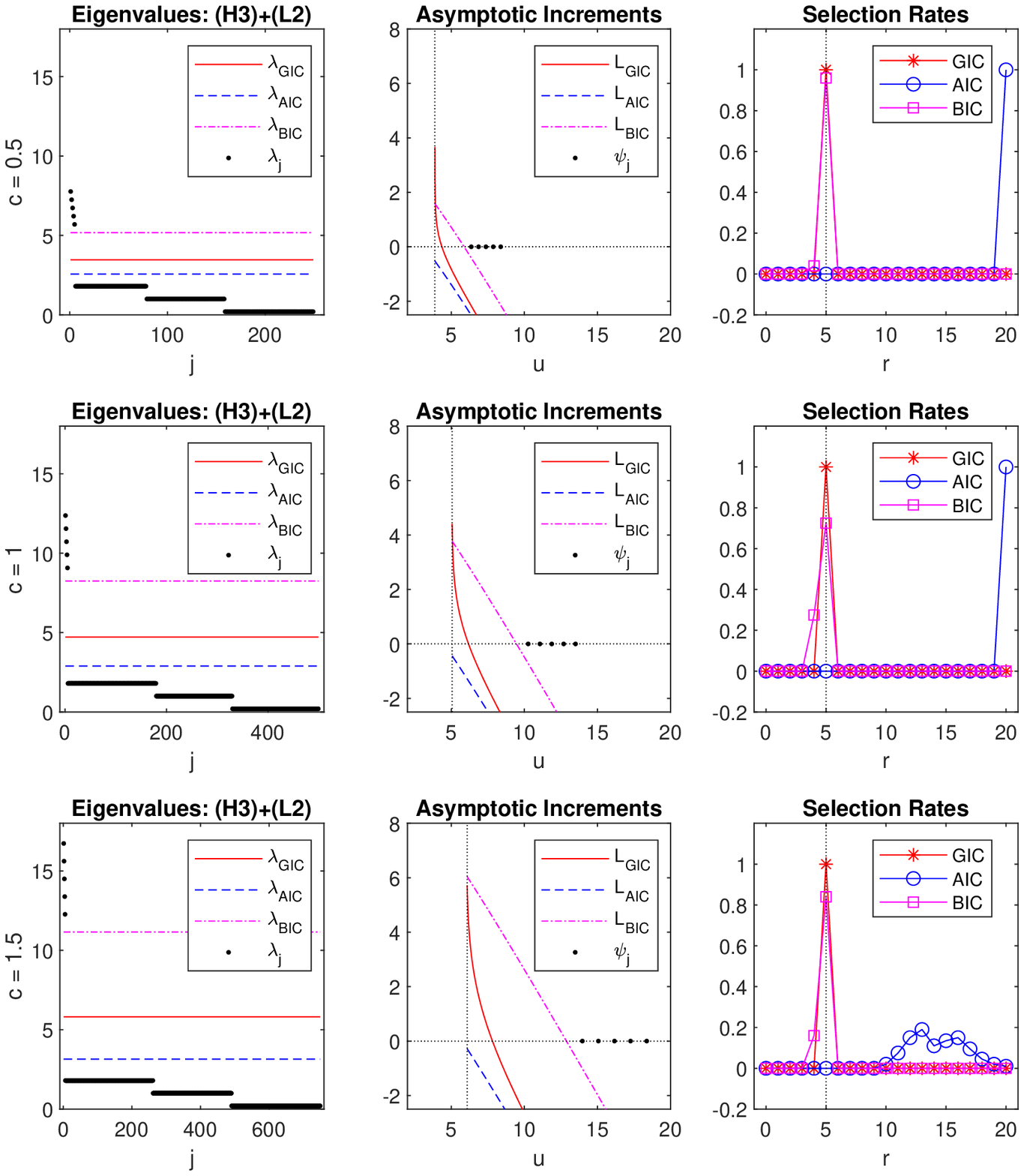}\\[-12ex]
\caption{Simulation results under (H3)+(L2) at different values of $c\in\{0.5,1,1.5\}$ (from top to bottom). The first column plots the true eigenvalues, where the horizontal lines represents $\lambda_\gic$ (the red solid line), $\lambda_\aic$ (the blue dashed line), and $\lambda_\bic$ (the magenta dash-dotted line). The second column plots the curves of $L_\gic$, $L_\aic$, and $L_\bic$, where the vertical dotted line represents the value of $b$, and the horizontal dotted line represents the value of 0. The values $\{\psi_j:j=1,\ldots,5\}$ of distant spikes are also reported. The third column reports the selection rate for each rank $r$, where the vertical dotted line indicates the true model size $r_0=5$.}
\label{fig.sim_H3_L2}
\end{figure}

Simulation results under (M2) are placed in Figures~\ref{fig.sim_H2_L1}--\ref{fig.sim_H2_L2}. In this case, $b$ has no closed form and is obtained  numerically. The selection consistency of GIC can still be observed from the third column of Figures~\ref{fig.sim_H2_L1}--\ref{fig.sim_H2_L2}, indicating the generality of GIC to different forms of~$H$. AIC is found to overestimate $r_0$ for $c\in\{0.5,1\}$, even for the case of larger signal eigenvalues (L2). This can be expected since the condition $L_\aic(b)>0$ of (A2) is violated for $c\in\{0.5,1\}$ (the second column of Figures~\ref{fig.sim_H2_L1}--\ref{fig.sim_H2_L2}). One situation to the success of AIC is the case with a large $c$, where (A2) can be easier to attain, and we observe the selection consistency of AIC for $c=1.5$. We remind the reader that it is difficult to determine a suitable size for $c$ in practical applications, since the form of $F_{c,H}$ is rarely known. Moreover, $c$ is not a tuning parameter but is determined from the data, which also limits the applicability of AIC. The inconsistency of AIC becomes more severe under (M3) as shown in Figures~\ref{fig.sim_H3_L1}--\ref{fig.sim_H3_L2}. In particular, condition $L_\aic(b)>0$ is found to be violated for all $c\in\{0.5,1,1.5\}$ (the second column of Figures~\ref{fig.sim_H3_L1}--\ref{fig.sim_H3_L2}). AIC largely overestimates the model size in all situations, while GIC is still found to achieve selection consistency (the third column of Figures~\ref{fig.sim_H3_L1}--\ref{fig.sim_H3_L2}). As for BIC, we still detect in (M2)--(M3) the same observation: BIC achieves selection consistency for the case of (L2) only. This observation supports our findings that BIC is generally less affected by the form of $H$, but its validity heavily relies on the extremely large signal eigenvalues (see the discussion below Theorem~\ref{corr.consistency}). The simulation results under (M2)--(M3) demonstrate the limitation of AIC and BIC and at the same time support the adaptivity of GIC to general $H$ with moderate size of signal eigenvalues.

We close this subsection by noting that the difference between GIC and conventional methods (AIC and BIC) can be clearly seen from the behavior of the curves $L_\gic$, $L_\aic$, and $L_\bic$. The $L$-curve, when associated with a good selection criterion, should have a sufficiently large value at $b$ so that the second gap condition, e.g., (G2) of GIC, can be satisfied, and it also should decay rapidly to 0 so that the first gap condition, e.g., (G1) of GIC, can be satisfied. For the cases of AIC and BIC, both $L_\aic$ and $L_\bic$ are nearly linear, and the two curves are parallel with a constant difference $(\ln n-2)c$. The reason is that both AIC and BIC adopt the penalty $b_r$ with a constant asymptotic increment $c$. This constant increment leads to the inflexibility of $b_r$-based selection criteria. When comparing AIC and BIC, they differ only in their different weights for $b_r$. Consequently, while $L_\bic(b)$ largely exceeds $L_\aic(b)$ (i.e., (B2) is easily satisfied), the linearity also causes the solution of $L_\bic(u)=0$ to largely exceed the solution of $L_\aic(u)=0$ at the same time (i.e., (B1) is easily violated). As for GIC, $L_\gic$ is a nonlinear function with a large value at~$b$ satisfying $L_\gic(b)\ge L_\aic(b)$, indicating that (G2) is more easily satisfied than (A2) of AIC. Moreover, $L_\gic$ decays more rapidly to 0 than BIC, indicating that (G1) is more easily satisfied than (B1) of BIC. These advantages of GIC come from its adaptivity to general eigenvalue structure and lead to the better applicability of GIC.

\subsection{Data analysis}

We illustrate the proposed GIC rank selection by applying it to the dataset analyzed in Wu et al. (2011),
which studies the habitual diet effect in the human gut microbiome. This dataset was also analyzed by Zheng, Lv and Lin (2020).
It consists of $n=91$ subjects, each with $301$ measurements (214 for nutrient intake and 87 for gut microbiome composition).
In our analysis, we first apply PCA to reduce the data dimension by preserving $99\%$ of the variation. This leads to a
resulting dimension $p=71$ of principal components. We then apply GIC, AIC, and BIC (with $q=20$) on the $91\times 71$ data matrix
(after componentwise standardization) to estimate the model rank.

\begin{figure}[!ht]
\hspace{-1.5cm}
\begin{center}
\includegraphics[width=6in]{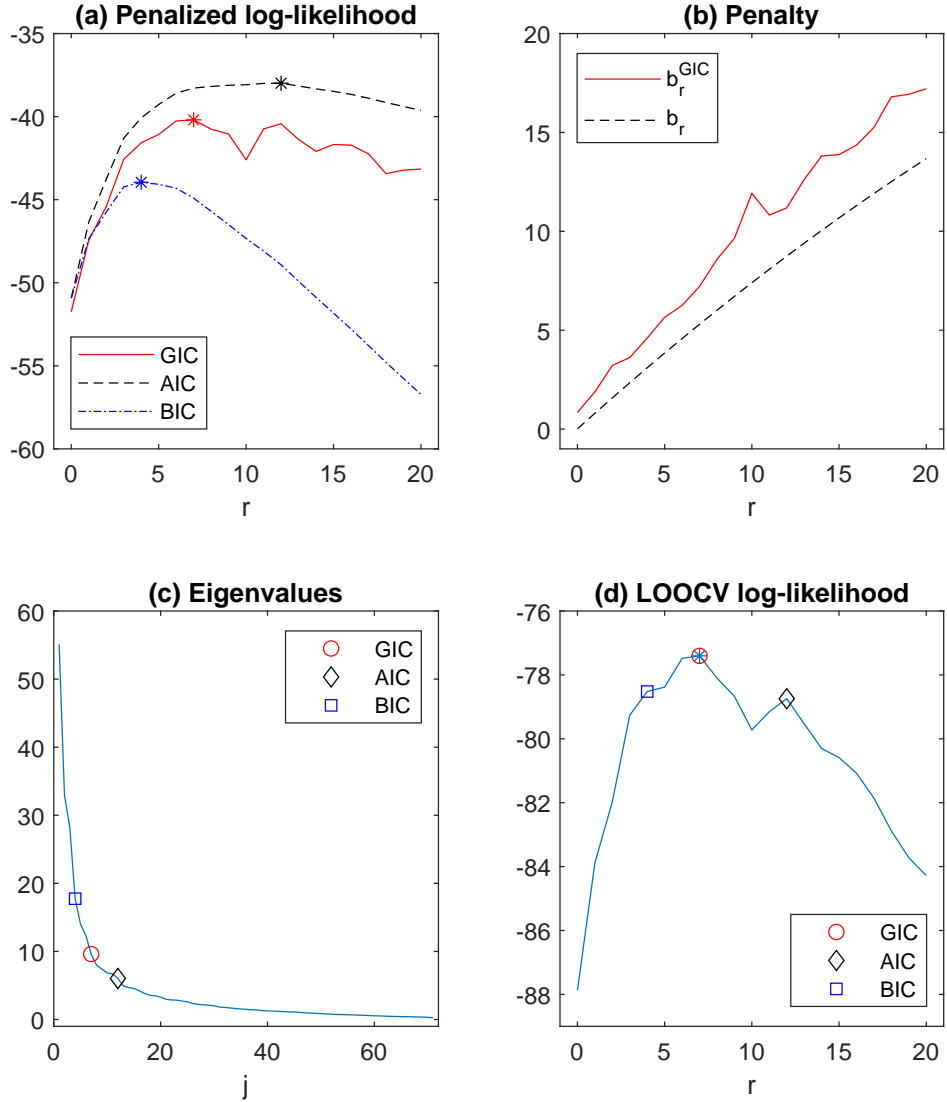}
\end{center}
\caption{Data analysis results. (a) Penalized log-likelihood from different criteria, where ``$*$'' indicates the maximum value for each method. These maximum values correspond to $\hat r_\bic =4$, $\hat r_\gic=7$, and $\hat r_\aic=12$. (b) Values of $\widehat b_r^\gic$ and $b_r$. (c) Eigenvalues of the sample covariance matrix, where $\hat\lambda_4$, $\hat\lambda_7$, and $\hat\lambda_{12}$ corresponding to the BIC, GIC, and AIC are indicated on the plot. (d) The values of LOOCV log-likelihood CV$(r)$, where ``$*$'' indicates the maximum value. The values at the estimated model rank from each method are also indicated.} \label{fig.data_analysis}
\end{figure}

The penalized log-likelihood of each method is presented in Figure~\ref{fig.data_analysis}~(a),
which gives $\widehat r_\gic=7$, $\widehat r_\aic=12$, and $\widehat r_\bic=4$. The penalty
functions $\widehat b_r^\gic$ and $b_r$ are presented in Figure~\ref{fig.data_analysis}~(b),
where $b_r$ is a smooth curve, while $\widehat b_r^\gic$ is not a monotone function, indicating
that $\widehat b_r^\gic$ is more adaptive to the underlying eigenvalue structures. Observe that $\widehat b_r^\gic$
has a peak at $r=10$ that also results in a local minimum of the penalized log-likelihood of GIC at $r=10$
(Figure~\ref{fig.data_analysis}~(a)). This sudden large value of $\widehat b_{10}^\gic$ indicates that the
values of the $10$th and $11$th eigenvalues are very close (see the eigenvalues in Figure~\ref{fig.data_analysis}~(c),
where $\widehat \lambda_9=7.4073$, $\widehat \lambda_{10}=6.8382$, $\widehat \lambda_{11}=6.7148$, and $\widehat \lambda_{12}=6.0254$).
By using $\widehat b_r^\gic$, the selection procedure tends to avoid an estimate of the model rank of $10$.
This scenario of nearly multiple eigenvalues for $\{\widehat \lambda_{10}, \widehat \lambda_{11}\}$
(which usually results in unstable analysis results if the model rank is selected at $10$), however,
is not detected by $b_r$, since it simply counts the number of free parameters without considering the eigenvalue
dispersion. To evaluate the estimated model rank from different methods, one can see in Figure~\ref{fig.data_analysis}~(c)
that an elbow cutoff looks around $r=8$ or $r=9$, which supports the result of $\widehat r_\gic=7$ better than $\widehat r_\aic=12$.
The estimate $\widehat r_\bic=4$ is too small, and useful information can be lost. However, the scree plot method can be
quite subjective and can depend on user bias. For a quantitative measure in evaluating the performance of
different selection criteria, we also report in Figure~\ref{fig.data_analysis}~(d) the leave-one-out cross-validated (LOOCV)
log-likelihood ${\rm CV}(r)=\frac{1}{n}\sum_{i=1}^{n}\ell(X_i|\widehat\theta_{r}^{(-i)})$ at $r\in\{0,1,2,\ldots,20\}$,
where $\widehat\theta_{r}^{(-i)}$ is the MLE of $\theta_r$ without using $X_i$. One can see that ${\rm CV}(r)$ achieves
the maximum at $r=7$, and the value ${\rm CV}(7)$ of GIC, is much larger than ${\rm CV}(12)$ of AIC and ${\rm CV}(4)$ of BIC. All the findings above support the selection result $\widehat r_\gic=7$.

\section{Application of GIC to Factor Analysis}

\subsection{Selecting the number of factors by GIC}

Factor analysis (FA) is a popular technique that is used to reduce the number of variables into a few factors. Unlike PCA, which aims to find uncorrelated variates in low dimension that accounts for most of the data variation, FA targets at latent factors that explain the variability among correlated covariates. FA assumes that $\Sigma$ takes the form
\begin{eqnarray}
\Sigma =L L^\top + D, \label{model_FA}
\end{eqnarray}
where $L$ is a $p\times q_0$ matrix of factor loading with $q_0<p$ and $D={\rm diag}(d_1,\ldots,d_p)$ consists of noise variances $\{d_j\}_{j=1}^p$. Note that $D$ is not necessarily proportional to $I_p$. The factor loading matrix $L$, which is the main interest of FA, is unique up to any orthogonal transformation multiplied on its right side. Similar to PCA, FA needs to determine the number of factors $q_0$. There are many methods developed for this problem, we will specifically focus on the parallel analysis (PA; Horn, 1965) based methods, which is later advanced by Buja and Eyuboglu (1992) and becomes one of the most popular methods for $q_0$ selection. PA is a sequential test based method, where the null distribution is generated by permutations. The result of PA thus depends on the realization of the permutations and the pre-specified significance level. Recently, Dobriban and Owen (2019) propose a deterministic version of PA (DPA) by using random matrix theory. They show that DPA and PA have similar performance in terms of selecting $q_0$, but DPA has some extra advantages of being reproducible,  not requiring to specify the significance level, and having low computational loading. They also further modify DPA and propose two variants of it, namely, the deflated DPA (DDPA) and the DDPA+. The latter uses a more accurate estimation of the singular value when doing deflation. The authors show that DDPA+ has the best performance in selecting $q_0$ among these DPA-based methods and thus recommend DDPA+ for being used in practice.

The number of factors $q_0$ in the FA model~(\ref{model_FA}) has certain connection to the target rank of $\Sigma$ (see Definition~\ref{r0}) that GIC aims to estimate. For example, when $D=\sigma^2 I_p$, (\ref{model_FA}) becomes the simple spiked covariance model. Then, $q_0$ is exactly the target rank of $\Sigma$, provided that the smallest eigenvalue of $LL^\top$ is larger than $\sqrt{c}\sigma^2$.
Dobriban (2017, Section~5.2) also discusses an application of PA to the problem of PCA rank selection, indicating the similarity between the factor selection of FA and the rank selection of PCA. These facts motivate us to apply GIC to selecting $q_0$ in FA, despite
finding sufficient conditions to ensure the equivalence of FA rank and PCA rank is beyond the scope of this article. In the next subsection, we will evaluate the performance of GIC in comparison with DPA-based methods via simulations.

\subsection{Numerical comparison of GIC with DPA-based methods}

We use the FA model~(\ref{model_FA}) to conduct simulations. The factor loading is set to be $L=\Xi\Phi$, where the columns of $\Xi\in \mathbb{R}^{p\times 3}$ represent the directions of factor loadings, and $\Phi\in \mathbb{R}^{3\times 3}$ is a diagonal matrix of signal sizes for the columns of $\Xi$. In our simulations, $\Xi$ is generated by normalizing the columns of a $p\times 3$ matrix with i.i.d. standard Gaussian entries for each simulation run. We consider the following four combinations for the distribution of $X$, the signal size $\Phi$, and the noise $D$:

\begin{itemize}
\item[(FA1)]
$X\sim N(0,\Sigma)$ with $\Phi=\sqrt{c}\cdot{\rm diag}(s, 10, 6)$ and $\{d_j\}_{j=1}^p\sim U(1,2)$. This is the setting used in Dobriban and Owen (2019).

\item[(FA2)]
$X\sim t_{30}(0,\Sigma)$ with $\Phi=\sqrt{c}\cdot{\rm diag}(s, 10, 6)$ and $\{d_j\}_{j=1}^p\sim U(1,2)$.
This setting is modified from (FA1) by using multivariate $t$-distribution for data generation.

\item[(FA3)]
$X\sim t_{30}(0,\Sigma)$ with $\Phi=\sqrt{c}\cdot{\rm diag}(s, 6, 6)$ and $\{d_j\}_{j=1}^p\sim U(1,2)$.
This setting is modified from (FA2) by considering the case of nearly multiple eigenvalues.
\item[(FA4)]
$X\sim t_{30}(0,\Sigma)$ with $\Phi=\sqrt{c}\cdot{\rm diag}(s, 6, 6)$ and $\{d_j\}_{j=1}^p\sim U(5,10)$.
This setting is modified from (FA3) by using more noisy data due to the larger diagonals in $D$.
\end{itemize}

\begin{figure}[!ht]
\hspace{-1cm}
\mbox{\includegraphics[width=3.6in]{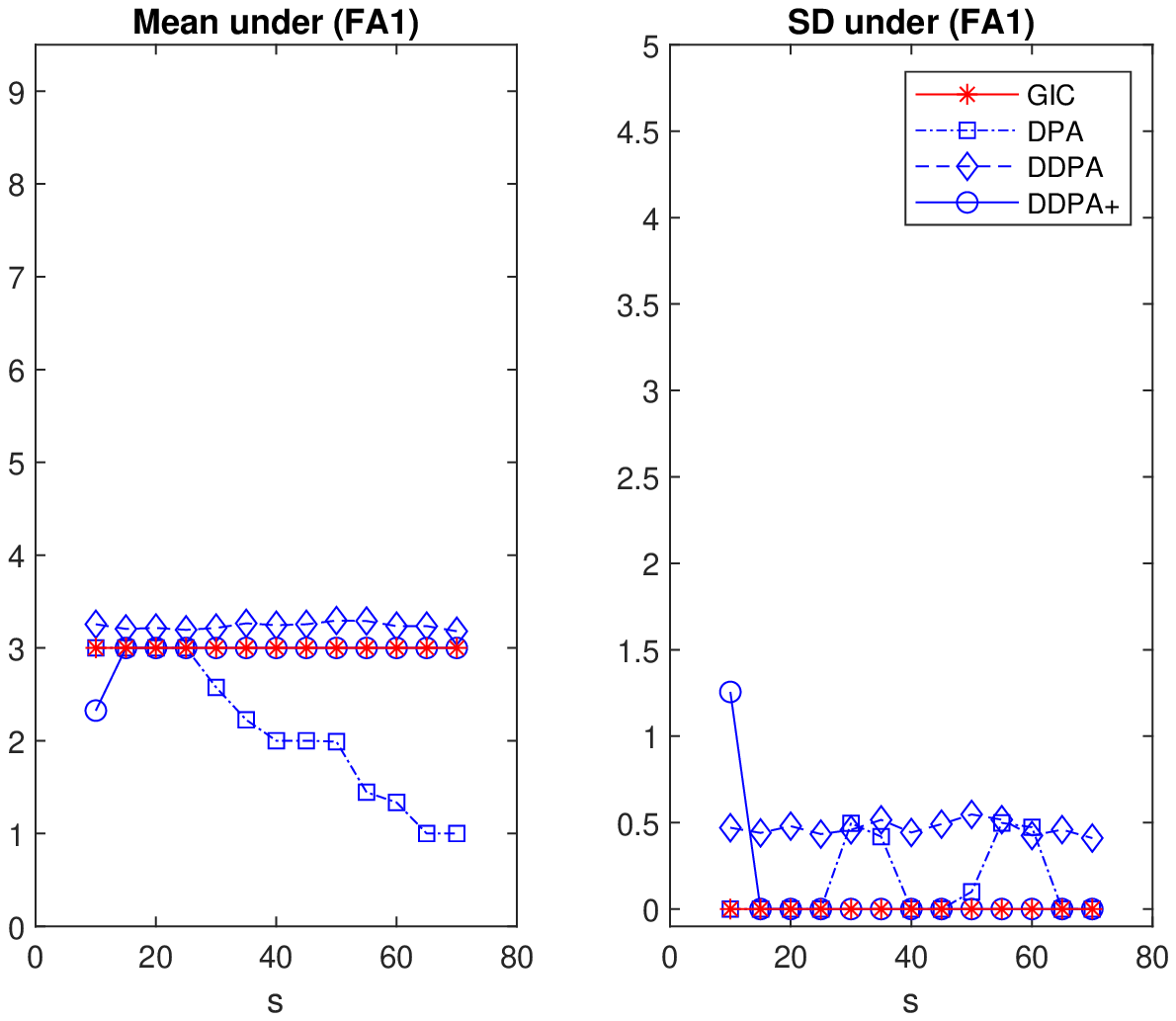}
\includegraphics[width=3.6in]{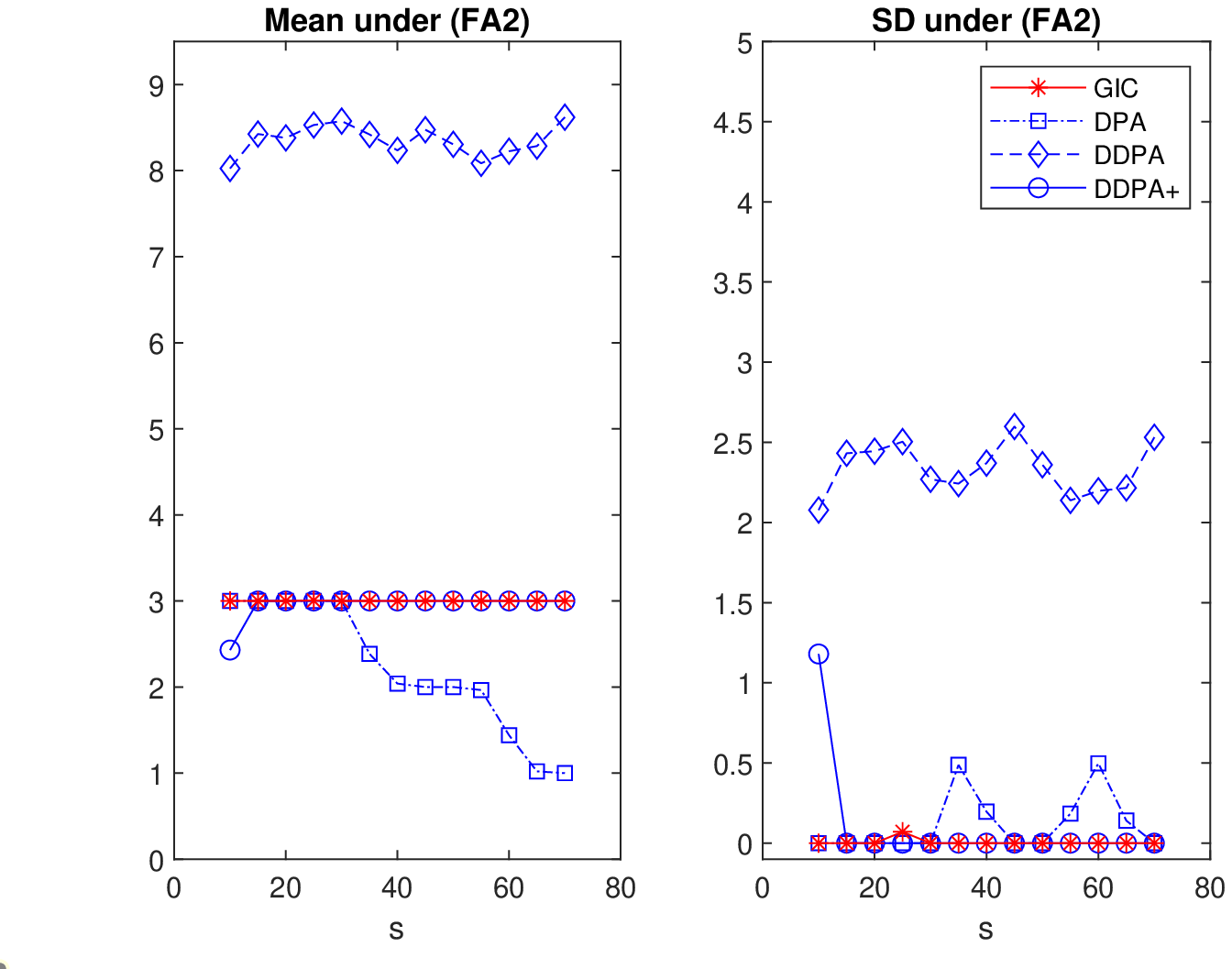}}

\hspace{-1cm}
\mbox{\includegraphics[width=3.6in]{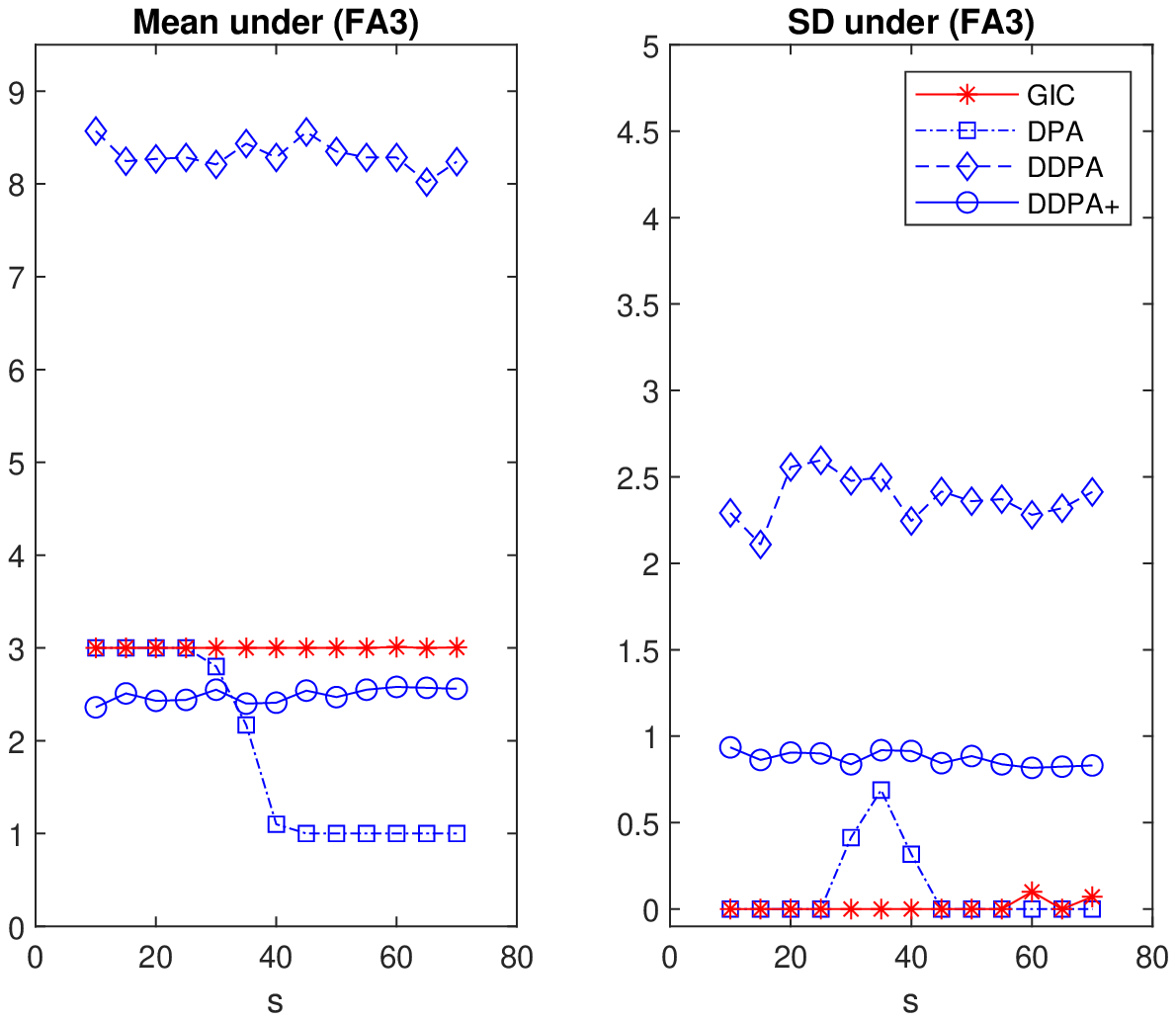}
\includegraphics[width=3.6in]{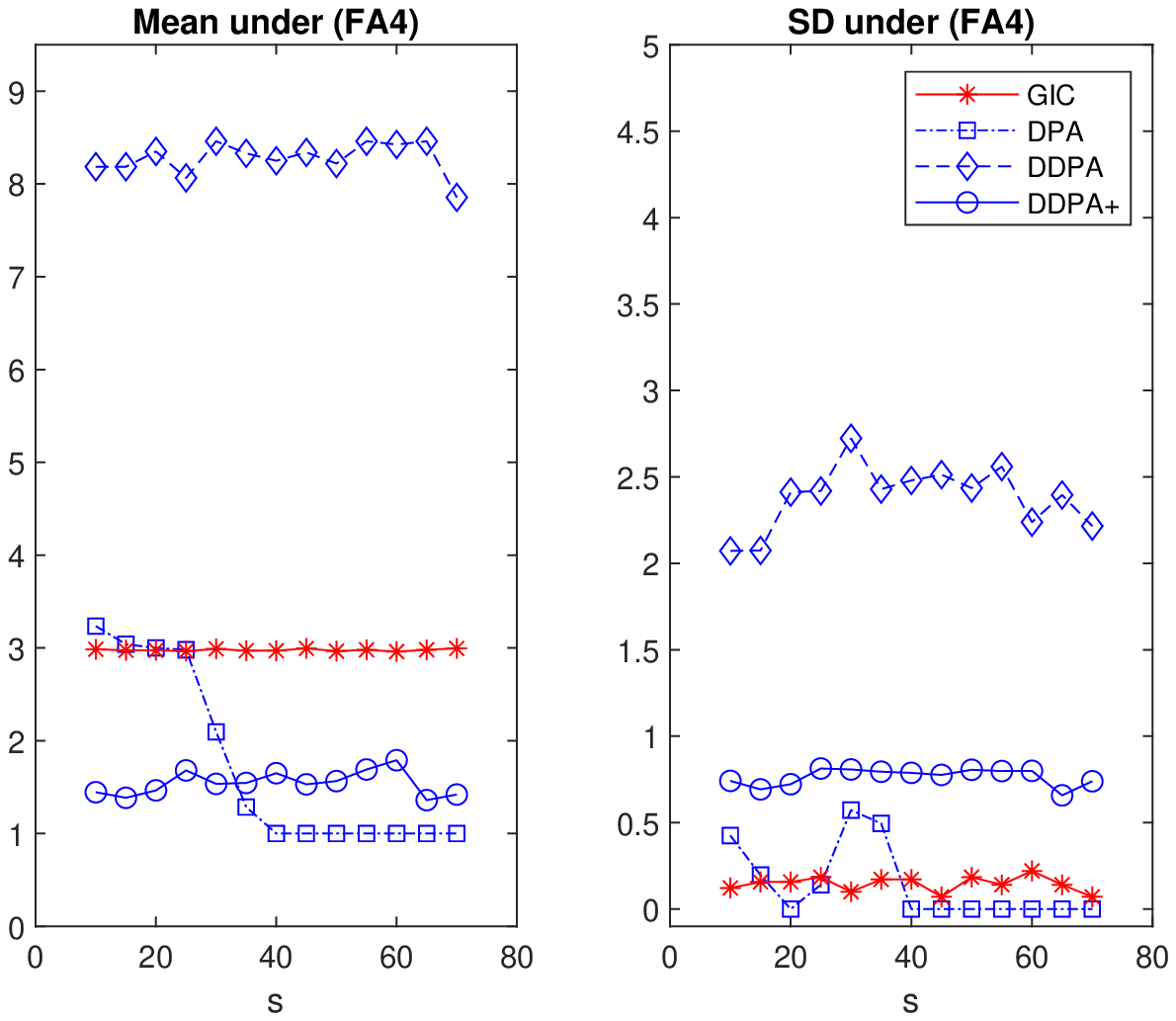}}
\caption{The means (Mean) and standard deviations (SD) of the selected number of factors from GIC, DPA, DDPA, and DDPA+ at different signal sizes $s$ under (FA1)-(FA4). The true number of factors is $q_0=3$ for all settings.} \label{fig.fa}
\end{figure}

\noindent In all settings, $(n,p)=(500,300)$, $c=0.6$, $q_0=3$, and $s\in\{10,15,20,\ldots,70\}$. For the implementation of DPA, DDPA and DDPA+, we use authors' code with default setting (\verb"github.com/dobriban/DPA"). As for PA (Buja and Eyuboglu, 1992), our simulation results indicate that it has similar performance as DPA, and is thus omitted from the comparison report.
The means and standard deviations (over 200 replicates) of the number of factors being selected by GIC (with $q=100$), DPA, DDPA, and DDPA+ are presented in Figure~\ref{fig.fa}.

For (FA1), it is obvious that GIC performs best among all methods. In particular, GIC consistently selects $q_0=3$ with small variance for all $s$. DPA is found to underestimate $q_0$ for cases with strong signal $s>30$, a phenomenon called ``shadowing'' by Dobriban (2017). DDPA does not suffer from ``shadowing'', but it overestimates $q_0$ for all $s$. DDPA+ performs well in selecting $q_0=3$,
except for the case of $s=10$, a case with nearly multiple eigenvalues, where a severe underestimation of $q_0$ is detected. The above mentioned weak points (shadowing in DPA, overestimation in DDPA, and underestimation in DDPA+ for the case with nearly multiple eigenvalues) are not observed in GIC. For (FA2), we see a pattern similar to (FA1) except that DDPA has an even more severe problem of overestimation. In (FA2), data are generated from $t_{30}$ instead of the Gaussian distribution. This implies that the deflation step of DDPA can be sensitive to the presence of outliers. Under (FA3), $\lambda_2\approx \lambda_3$ for all $s$, and the problem of underestimation of DDPA+ is detected for all $s$. DDPA+ is also found to have larger standard deviation than GIC. On the other hand,
recall that GIC is less affected by the presence of multiple eigenvalues, which has been discussed earlier in Section~\ref{sec.derivation_gic}. This robustness is also supported by the empirical results under (FA3), where GIC consistently selects $q_0=3$ for all $s$. For the more noisy case (FA4), the problem of underestimation of DDPA+ becomes even more severe. DPA is also found to have selection bias for $s\le 30$, despite it performs well for $s\le 30$ under (FA1)-(FA3), indicating that DPA can be sensitive to higher noise level. As to GIC, we still observe its consistency for all $s$, except a slightly larger variation is detected.

In conclusion, our numerical studies demonstrate the superiority of GIC over DPA-based methods in selecting the number of factors. The superiority of GIC comes from its construction under the generalized spiked covariance model, and GIC is found to be robust to the signal size, to the underlying data generating distribution, to the presence of multiple eigenvalues, and to the noise levels.

\begin{rmk}
Like DDPA+, GIC is also a deterministic method without involving random permutations or the specification of tuning parameters. The computational burden of GIC is one single SVD for $S_n$ only. We thus omit the comparison of computational time for GIC with DDPA+.
\end{rmk}

\section{Concluding Discussions}

A novel GIC-based selection criterion is proposed to determine the rank of the PCA model.
Unlike the conventional AIC and BIC, which simply count the number of free parameters, GIC takes into account the
sizes of $\lambda_j$'s and is hence more adaptive to various structures of population covariance. The performance of GIA, AIC, and BIC is affected by $(c,H)$ and the size of $\psi_{r_0}/\mu_H$, and each method has its own merit under different scenarios:
\begin{itemize}
\item
AIC requires the smallest size of $\psi_{r_0}/\mu_H$ among the three methods. The cost,
however, is that AIC is very sensitive to deviations in $H$ from the simple spiked covariance model, and is not
suitable for the case of a small to moderate $c$.

\item
BIC is the most robust method among the three to various forms of $H$, and to the size of $c$. The cost is that the selection consistency of BIC requires the largest size $\psi_{r_0}/\mu_H$ among the three methods,
i.e., BIC only identifies directions with strong signals.

\item
GIC adapts to deviations in $H$ from the simple spiked covariance model, and to the size of $c$. Moreover, the required
size of $\psi_{r_0}/\mu_H$ for the selection consistency of GIC is between AIC and BIC, and does not diverge with $\ln n$ as does BIC.
\end{itemize}

\noindent In conclusion, there is no definite choice among the three criteria GIC, AIC, and BIC. Through the theoretical investigation of GIC, we have provided more insights for different behaviors
of the three methods. In particular, GIC tends to select a model of a size between AIC and BIC (see Remark~\ref{rmk.bic_gic_aic}),
and is a more robust selection criterion (against the form of $H$, the size of $c$, and the sizes of distant spike eigenvalues) for practical applications.

We have assumed $\kappa(b/\mu_H)<\infty$, and under this assumption the asymptotic increment of GIC is shown to be $\kappa_j$ as discussed in Section~\ref{sec.consistency_GIC}. Also under this assumption the gap conditions are derived to ensure the selection consistency of GIC.
The existence of finite $\kappa(b/\mu_H)$, however, depends on the characteristic of $H$. We have shown that $\kappa(b/\mu_H)=2c+c\sqrt{c}<\infty$ under the simple spiked model assumption (C5), but we cannot extend this result to general $H$, due to the lack of the explicit form of $F_{c,H}$. We would like to emphasize whether or not the finiteness of $\kappa(b/\mu_H)$ will not hinder one from using the selection rule $\widehat r_\gic$ because the expression for $b_r^\gic$ remains unchanged. However, the gap conditions under diverging $\kappa(b/\mu_H)$ need to be modified to achieve selection consistency. Since this modification deserves separate treatment, it is not further pursued here.

Finally, the validity of applying GIC to selecting the number of factors in FA is based on the assumption that $q_0$ in~(\ref{model_FA}) is equivalent to the target rank of $\Sigma$, which is rigorously defined in Definition~\ref{r0}. For the simple spiked covariance model, $q_0$ in FA coincides with the target rank in PCA. For more general cases beyond the simple spiked  covariance model, it is of interest to have further investigations and to derive GIC specifically for the FA model~(\ref{model_FA}).

\section*{References}
\begin{description}
\item
Absil, P.A.,  Mahony R. and Sepulchre, R. (2008). {\it Optimization Algorithms on Matrix Manifolds}. Princeton University Press.

\item
Bai, Z.D., Choi, K.P. and Fujikoshi, Y. (2018).
Consistency of AIC and BIC in estimating the number of
significant components in high-dimensional principal component analysis. {\it
Annals of Statistics}, 46, 1050--1076.

\item
Bai, Z., Chen, J., and Yao, J. (2010). On estimation of the
population spectral distribution from a high-dimensional sample
covariance matrix. {\it Australian and New Zealand Journal of
Statistics}, 52, 423--437.

\item
Bai, Z.D. and Silverstein, J.W. (2010). {\it Spectral Analysis
of Large Dimensional Random Matrices}, 2nd ed., Springer.

\item
Bai, Z.D. and Yao, J. (2012). On sample eigenvalues in a generalized
spiked population model. {\it Journal of Multivariate Analysis}, 106,
167--177.

\item
Bai, Z.D., Yin, Y.Q. and Krishnaiah, P.R. (1987). On the limiting empirical distribution function of the eigenvalues of a multivariate $F$ matrix. {\it Theory of Probability and its Applications}, 32, 537--548.

\item
Beutner, E. and Z\"ahle, H. (2010). A modified functional delta method and its application to the estimation of risk functionals. {\it Journal of Multivariate Analysis}, 101, 2452--2463.

\item
Buja, A., and Eyuboglu, N. (1992). Remarks on parallel analysis. {\it Multivariate behavioral research}, 27, 509-540.

\item
Critchley, F. (1985). Influence in principal component analysis. {\it Biometrika}, 72, 627--636.

\item
Dobriban, E. (2017). Factor selection by permutation. arXiv:1710.00479.

\item
Dobriban, E., and Owen, A. B. (2019). Deterministic parallel analysis: an improved method for selecting factors and principal components. {\it Journal of the Royal Statistical Society: Series B}, 81, 163-183.

\item
Findley, D. F., and Wei, C. Z. (2002). AIC, overfitting principles, and the boundedness
of moments of inverse matrices for vector autoregressions and related models. {\it Journal of Multivariate Analysis}, 83, 415--450.

\item
Fujikoshi, Y., and Sakurai, T. (2016). Some Properties of Estimation Criteria for Dimensionality in Principal Component Analysis. {\it American Journal of Mathematical and Management Sciences}, 35, 133-142.

\item
Horn, J. L. (1965). A rationale and test for the number of factors in factor analysis. {\it Psychometrika}, 30, 179-185.

\item
Hsu, H. L., Ing, C. K., and Tong, H. (2019).
On model selection from a finite family of possibly misspecified time series models.
{\it Annals of Statistics}, 47, 1061--1087.

\item
Konishi, S. and Kitagawa, G. (1996). Generalized information
criteria in model selection. {\it Biometrika}, 83, 875--890.

\item
Lv, J. and Liu, J. S. (2014). Model selection principles in misspecified models.
{\it Journal of the Royal Statistical Society: Series B (Statistical Methodology)},
76, 141--167.


\item
Wu, G. D., Chen, J., Hoffmann, C., Bittinger, K., Chen, Y. Y., Keilbaugh, S. A., Bewtra, M., Knights, D., Walters, W. A., Knight, R.,
Sinha, R., Gilroy, E., Gupta, K., Baldassano, R., Nessel, L., Li, H., Bushman, F. D., and Lewis, J. D. (2011).
Linking long-term dietary patterns with gut microbial enterotypes. {\it Science}, 334, 105--108.


\item
Yao, J., Zheng, S. and Bai, Z. (2015).
{\it Large Sample Covariance Matrices and High-Dimensional Data Analysis}. New York, NY: Cambridge University Press.

\item
Zheng, Z., Lv, J. and Lin, W. (2020). Nonsparse learning with latent variables. {\it Operations Research}, to appear.
\end{description}

\newpage

\end{document}